\documentclass[12pt,a4paper,fleqn,usenatbib]{mnras}
\usepackage{newtxtext,newtxmath}
\usepackage[T1]{fontenc}
\usepackage{ae,aecompl}
\usepackage{graphics}
\usepackage{graphicx}
\usepackage{amsmath}	
\usepackage[flushleft]{threeparttable}
\usepackage{booktabs,caption}
\usepackage{placeins}
\usepackage{float}

\newcommand{\unitfl}{$\text{erg}/\text{s}/\text{cm}^{2}\,$}
\newcommand{\unilum}{$\text{erg}/\text{s}\,$}

\title[MAXI J0655-013]{Study of recently discovered Be/X-ray pulsar MAXI J0655-013 using \emph{NuSTAR}}
\author[Rai et al.]
{Binay Rai,$^{1}$\thanks{binayrai21@gmail.com}
Mohammed Tobrej,$^{1}$\thanks{tabrez.md565@gmail.com} 
Manoj Ghising,$^{1}$\thanks{manojghising26@gmail.com}
Ruchi Tamang,$^{1}$\thanks{ruchitamang76@gmail.com}
\newauthor
Bikash Chandra Paul$^{1}$\thanks{bcpaul@nbu.ac.in;bcpaul@associates.iucaa.in}
\\
$^{1}$Department of Physics, North Bengal University,Siliguri, Darjeeling, WB, 734013, India
\\
}

\pubyear{2022}

\begin{document}
\label{firstpage}
\pagerange{\pageref{firstpage}--\pageref{lastpage}}
\maketitle
\begin{abstract}
We study the recently discovered Be/X-ray pulsar MAXI J0655--013 using the 2022 \emph{NuSTAR} observations. The paper is the first detailed study of the timing and spectral properties of the source. The pulse profiles of the pulsar vary with energy. The pulsed fraction is found to increase monotonically with energy. In between the two \emph{NuSTAR} observations, a large spin-up rate of $\sim$ -1.23 s d$^{-1}$ is observed, which can be due to large spin-up torque acting on the pulsar during an outburst. Such a large spin-up rate is observed for the first time in an X-ray pulsar during an outburst. The variation of the spin period with time can be employed to obtain the orbital parameters of the binary system, and we found the orbital period to be $\sim$ 27.9 d. The second \emph{NuSTAR} observation is done in a low luminosity state ($L_{X} \sim$3.9$\times$10$^{34}$ \unilum). We have detected the pulsation of the pulsar in such a low luminosity state. In such a low luminosity state, the pulsar MAXI J0655--013 might be accreting from the cold disk.
\end{abstract}


%
%
\section{Introduction}

MAXI J0655--013 is an X-ray transient first detected by \emph{MAXI} at 10:23 UTC on June 18, 2022 \citep{2022ATel15442....1S}. The coordinates of the source estimated by \emph{MAXI} were $RA=06^{h}55^{m}27^{s}$ and $Dec.=-01^{\circ}23^{\prime}52^{\prime\prime}$. The average X-ray flux estimated for \emph{MAXI} in the 4–10 keV energy range is 61$\pm$8 mCrab \citep{2022ATel15442....1S}. On the same day at 20:26:02 UTC, the \emph{Swift}-BAT was triggered along the direction of the source \citep{2022ATel15443....1K} and was confirmed to be a new transient X-ray source. The positional coordinates of the source provided by \emph{Swift} were $RA=06^{h}55^{m}10^{s}$, $Dec.=-01^{\circ}31^{\prime}49^{\prime\prime}$ \citep{2022ATel15443....1K}. An increase in flux on the next day of detection was reported by \cite{2022ATel15453....1N}. The exact coordinates of the source were measured by \cite{2022ATel15564....1K} and they were $RA=06^{h}55^{m}12.37^{s}$ and $Dec=-01^{\circ}28^{\prime}52.7^{\prime\prime}$.
A ToO observation was performed by \emph{NuSTAR} between June 21-22, 2022. A pulsation of 1129.1$\pm$0.3 s was detected during the \emph{NuSTAR} \citep{2022ATel15495....1S}, thus confirming the source to be an X-ray pulsar. The flux measured in the energy range 3-50 keV from \emph{NuSTAR} observations is 3.3$\times$10$^{-9}$ \unitfl \citep{2022ATel15495....1S}. The source position estimated by the \emph{NuSTAR} was 1$^{\prime}$ apart from that of another X-ray source, 2SXPS J065512.4-012855 (aka SRGA J065513.5-012846), whose optical counterpart is a Be star V520 Mon and located at a distance of 4.1 kpc \citep{2022A&A...661A..38P}. \cite{2022ATel15561....1K} confirmed that V520 Mon is an optical counterpart of MAXI J0655--013, and the source is the same as 2SXPS J065512.4-012855 observed by \emph{Swift} earlier on 23$^{rd}$ of May 2018. The optical spectrum of the V520 Mon shows a prominent H$_{\alpha}$ line, and the preliminary result suggests that it is a B1-3e III-V star \citep{2022ATel15582....1Z}. Thus, the source MAXI J0655--013 is a Be/X-ray pulsar. \cite{2022ATel15612....1R} using the optical observation of V520 Mon from Skinakas observatory added further evidence that it is a Be/X-ray binary (BeXRB). \cite{2022ATel15612....1R} observed the spectral type of the companion to be 0.95-B0V.
BeXRB is a class of neutron stars High Mass X-ray binary (NS HMXB). Most of the BeXRBs are transient and observable only during bright outbursts. Although a few persistent BeXRBs are also observed \citep{2011Ap&SS.332....1R}. Persistent BeXRBs are characterised by very low luminosity ($L_{X}\lesssim10^{35}$ \unilum), less X-ray variability, a slow rotational period ($P_{s}>200$ s), and they do not undergo bright outbursts. Two types of outbursts are observed in BeXRBs, namely, Type I and Type II. A Type I outburst is nearly periodic and occurs during periastron passage when the neutron star passes through the circumstellar disc around the Be star. The duration of a typical Type I outburst lies between 0.2-0.3 P$_{orb}$ \citep{2011Ap&SS.332....1R} with a peak luminosity of $\le$ 10$^{37}$\unilum. On the other hand, Type II outbursts are very bright, and the corresponding peak luminosity may reach above the Eddington limit ($\ge$ 10$^{38}$ \unilum). This type of outburst is non-periodic and may last for a few orbital periods. \cite{Okazaki2013OriginOT} proposed a scenario where a misaligned and warped Be circumstellar disk crosses the neutron star's orbit. Thereafter, the neutron star captures a huge amount of matter through the Bondi-Hoyle-Lyttleton (BHL) process.

\section{Observations and Data Reduction}
\subsection{NuSTAR}

\begin{figure*}
\begin{minipage}{0.33\textwidth}
\includegraphics[width=9 cm, height=7cm]{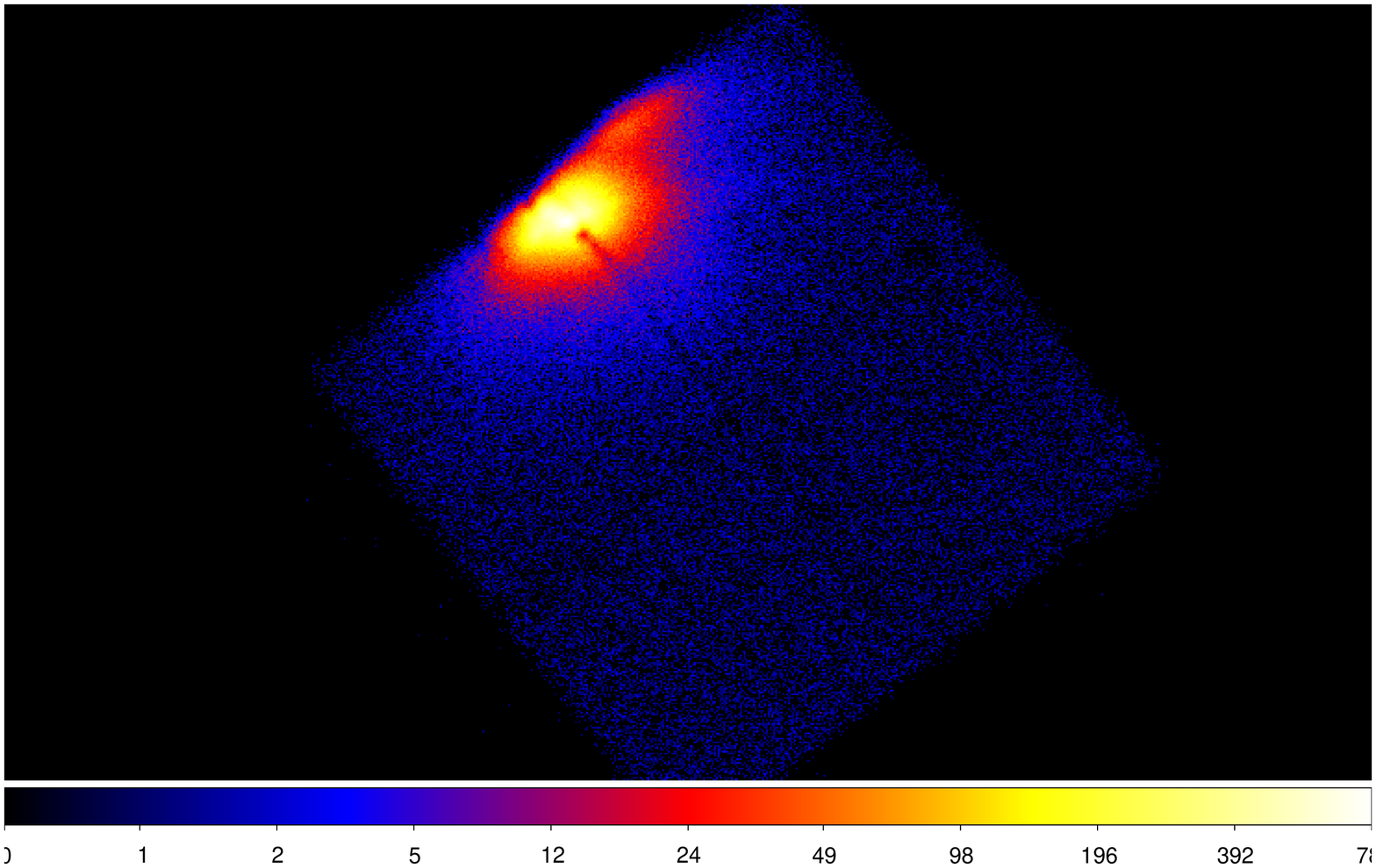}
\end{minipage}
\hspace{0.2\linewidth}
\begin{minipage}{0.33\textwidth}
\includegraphics[width=8.5 cm, height=7cm]{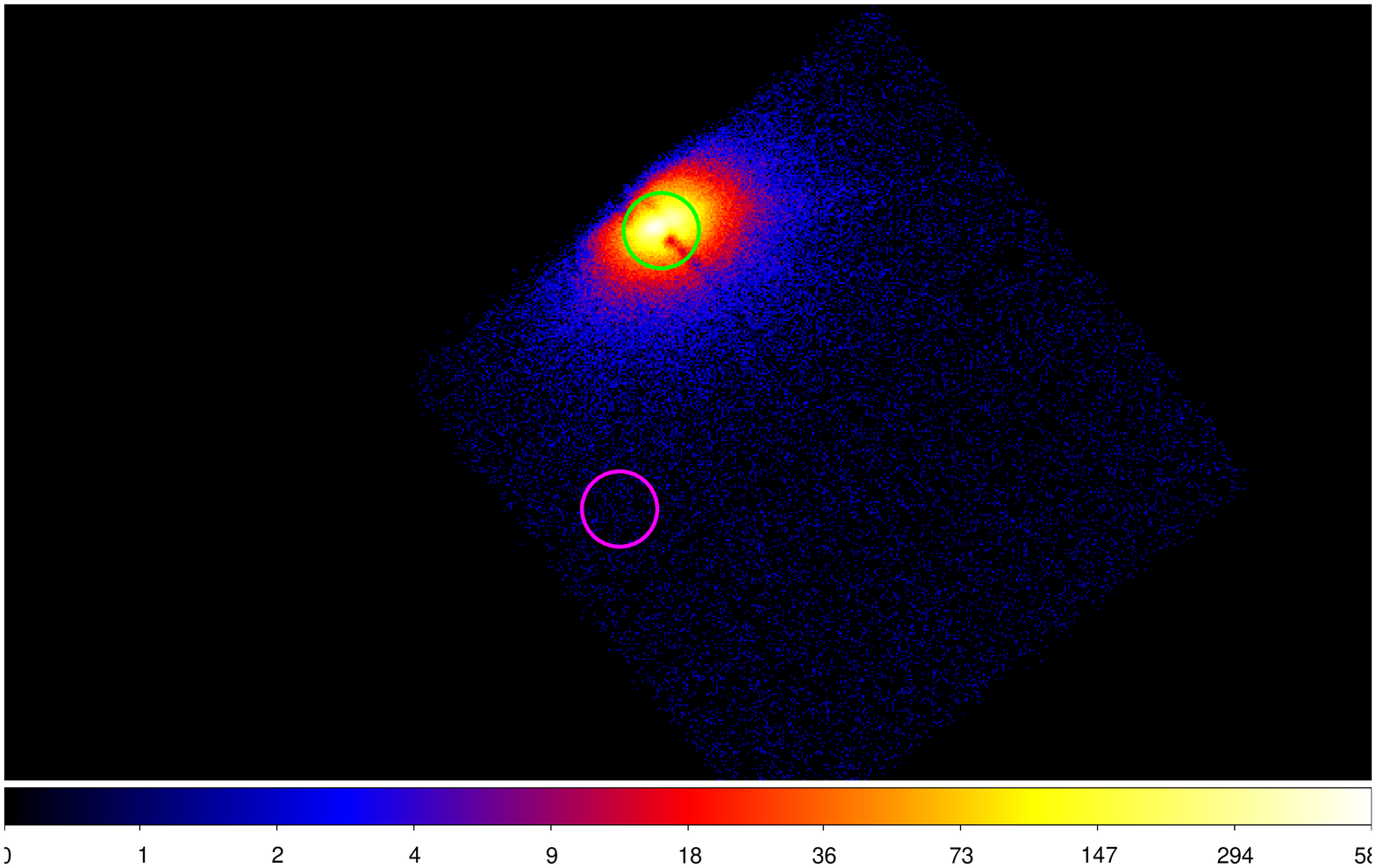}
\end{minipage}
\caption{\textit{Left}- The image of the source obtained from the combinations of all CHUs (CHU1, CHU2 \& CHU3) in the absence of the primary Camera Header Unit (CHU4) for the \emph{NuSTAR} FPMA. \textit{Right} The image of the pulsar MAXI J0655-013 obtained by \emph{NuSTAR's} CHU13 for the same instrument. The green and magenta circles indicates the source and background extraction regions respectively.}
\label{fig1}
\end{figure*}

A \textsc{heasoft} \footnote{\url{https://heasarc.gsfc.nasa.gov/docs/software/heasoft/download.html}} (v6.30.1) software was used for data reduction and analysis. A NASA space-based X-ray telescope, \texttt{Nuclear Spectroscopic Telescope Array} (\emph{NuSTAR}), also known as SMEX-11 and Explorer 93, is employed to examine high-energy X-rays originating from astrophysical sources. \emph{NuSTAR} is the first hard X-ray focusing telescope that works in the energy band of 3-79 keV, it is made up of two identical X-ray telescope modules with separated focal plane modules, namely, FPMA and FPMB \citep{Harrison_2013}. The spectral resolution of the telescope at 10 keV is 400 eV and has an angular precision of 18 $^{\prime\prime}$. It has a temporal resolution of 2 $\mu$s and is suitable for timing analysis. The \emph{NuSTAR} data of the MAXI J0655-013 were processed using mission-specific software \textsc{nustardas} v2.1.1 along with the latest calibration database v20220829. The standard data screening and filtering tool \textsc{nupipeline} was used to obtain the clean event files for the two modules. 
The \emph{NuSTAR} observed the source twice (see Figure \ref{fig2}), one at the beginning of the outburst and the other after the outburst. The observation IDs of the \emph{NuSTAR} observations are 80801347002 and 90801321001, and hereafter we refer to two observations as Obs1 and Obs2 respectively (see Table \ref{Tab1}). An onboard star tracker resting on the optical bench (Camera Head Unit $\#$4, CHU4) of the \emph{NuSTAR} is primarily used for determining the telescope's absolute pointing, which is then used for the reconstruction of the sky coordinates in the cleaned event files of Mode 01. The Mode 01 cleaned event files are primarily used for scientific study. However, in the case of Obs1, CHU4 was possibly blinded by a bright source or Earth \footnote{\url{https://heasarc.gsfc.nasa.gov/docs/nustar/analysis/nustar_swguide.pdf}} \citep{2022ApJ...927..190P}, due to which it was not available. In the absence of a primary star tracker, the aspect reconstruction is done using the three-star trackers (CHU1, CHU2 \& CHU3) located on the spacecraft bus. In this case, a cleaned event file of Mode 06 is used for scientific study \citep{2022ApJ...927..190P}. The reconstruction of the source image in Mode 06 (see left of Figure \ref{fig1}) is not perfect, as a result of which one can observe multiple centroids in the source image. Each centroid corresponds to a particular CHUs combination. However, we can split the Mode 06 cleaned event file in terms of different CHUs' combinations using the tool \textsc{nusplitsc}. Using the tool, we have split the Mode 06 cleaned event files in strict splitting mode into different cleaned event files depending on the CHUs' combinations. A source image obtained from CHU13 is shown in the right of Figure \ref{fig1}.

In the case of Obs1, the source was found to lie in the gap between the two detectors and was lying near the edge of the detectors. We have used DS9 \citep{2003ASPC..295..489J} for selecting the source and background extraction regions for each CHU combination. A circular region of radius 50$^{\prime\prime}$ along the source was considered the source extraction region. The centre of the source was detected using the DS9 automatic centroid determination algorithm. Another circular region of the same radius in the source-free region was taken as the background extraction region. The exposure time of CHU1 was very short ($<$100 s) due to which we have not considered cleaned event files from CHU1. We chose a background region from the same detector on which the centroid falls. For each CHUs combination, the source and background spectra and light curves were then generated using \textsc{nuproducts}. In case of Obs2, too, we took a circular region of 50$^{\prime\prime}$ radius centered around the source and considered it the source extraction region. A background extraction region of the same size was considered free from the source. A tool \textsc{lcmath} was used for background subtraction from the light curves. The background-subtracted light curve was then transformed to barycenter using \textsc{barycorr}.

\begin{table}
\begin{tabular}{c c c}
\hline
Obs ID	&	Exposure (ks)	&		Start time of the Observation \\
\hline	
80801347002$^{\ast}$		&	\dots	&   2022-06-21 22:31:09 \\
90801321001	&	11.3	&		2022-08-14 20:06:09	\\
\hline
\end{tabular}
\caption{Details about the \emph{NuSTAR} observation of MAXI J0655-013. $^{\ast}$The different CHUs combinations had different exposure in this case - CHU2, CHU12, CHU13, CHU23, and CHU123 had an exposure of 17.1 ks, 6.3 ks, 12.1 ks, 3 ks, and 1.2 ks respectively.}
\label{Tab1}
\end{table}

\section{Results}
\subsection{Light curve}

\begin{figure}
\includegraphics[scale=0.35]{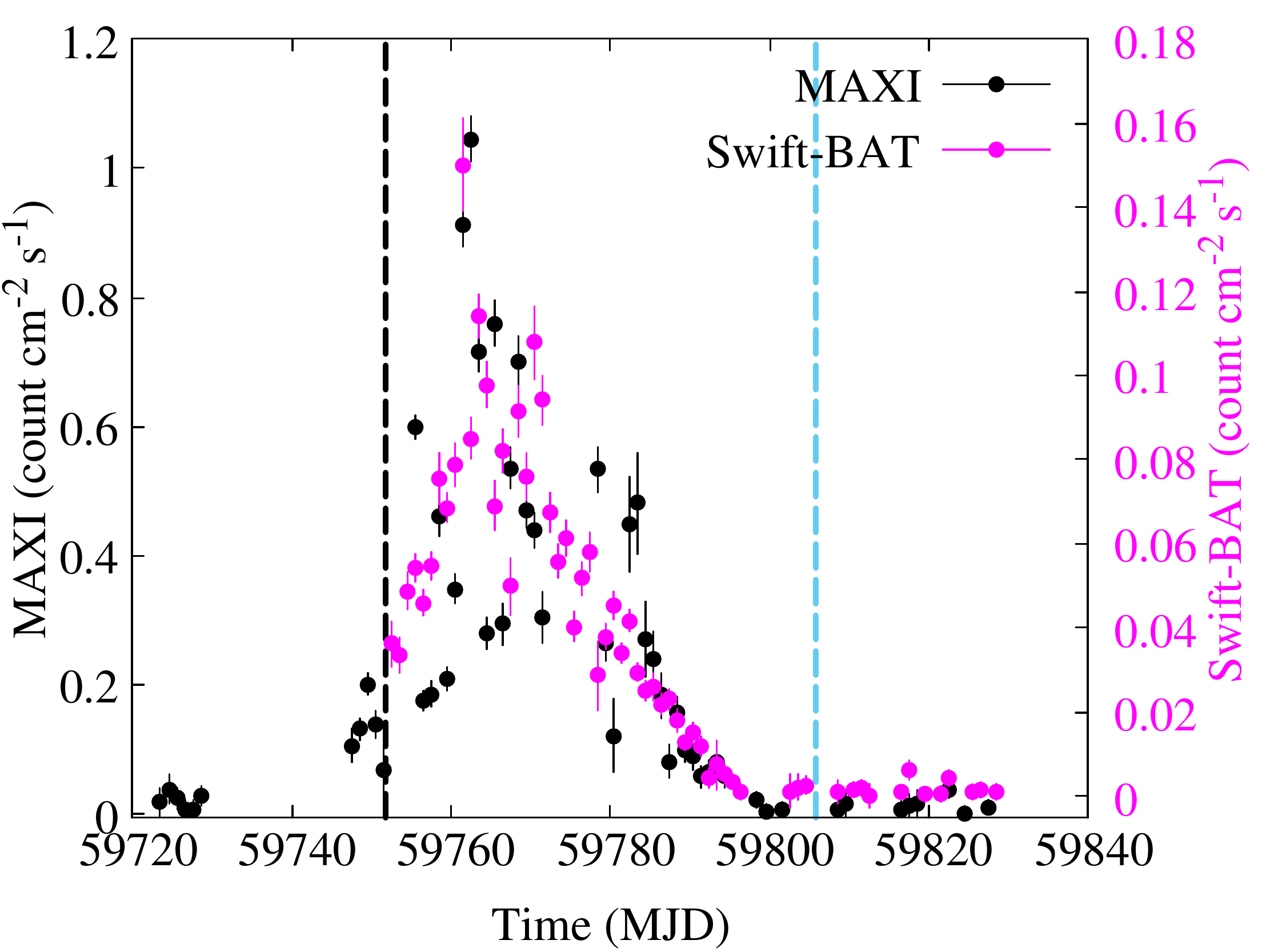}
\caption{Light curve of MAXI J0655-013, the black and magenta coloured points indicates \emph{MAXI} (2-20 keV) and \emph{Swift}-BAT (15-50 keV) count rates respectively. The dashed line in black and cyan indicates the \emph{NuSTAR's} Obs1 and Obs2 respectively.}
\label{fig2}
\end{figure}

The light curves of the source during outburst are plotted in Figure \ref{fig2}. The time coverage of the \emph{MAXI} was more than \emph{Swift}-BAT, the former did not cover the rising phase of the outburst. The duration of the outburst was $\sim$ 52 days. As evident from Figure \ref{fig2}, the Obs1 was done in the initial phase of the outburst and the Obs2 taken during the quiescent phase after the outburst.

\subsection{Pulse period}

\begin{figure*}
\begin{minipage}{0.30\textwidth}
\includegraphics[height=1.1\columnwidth]{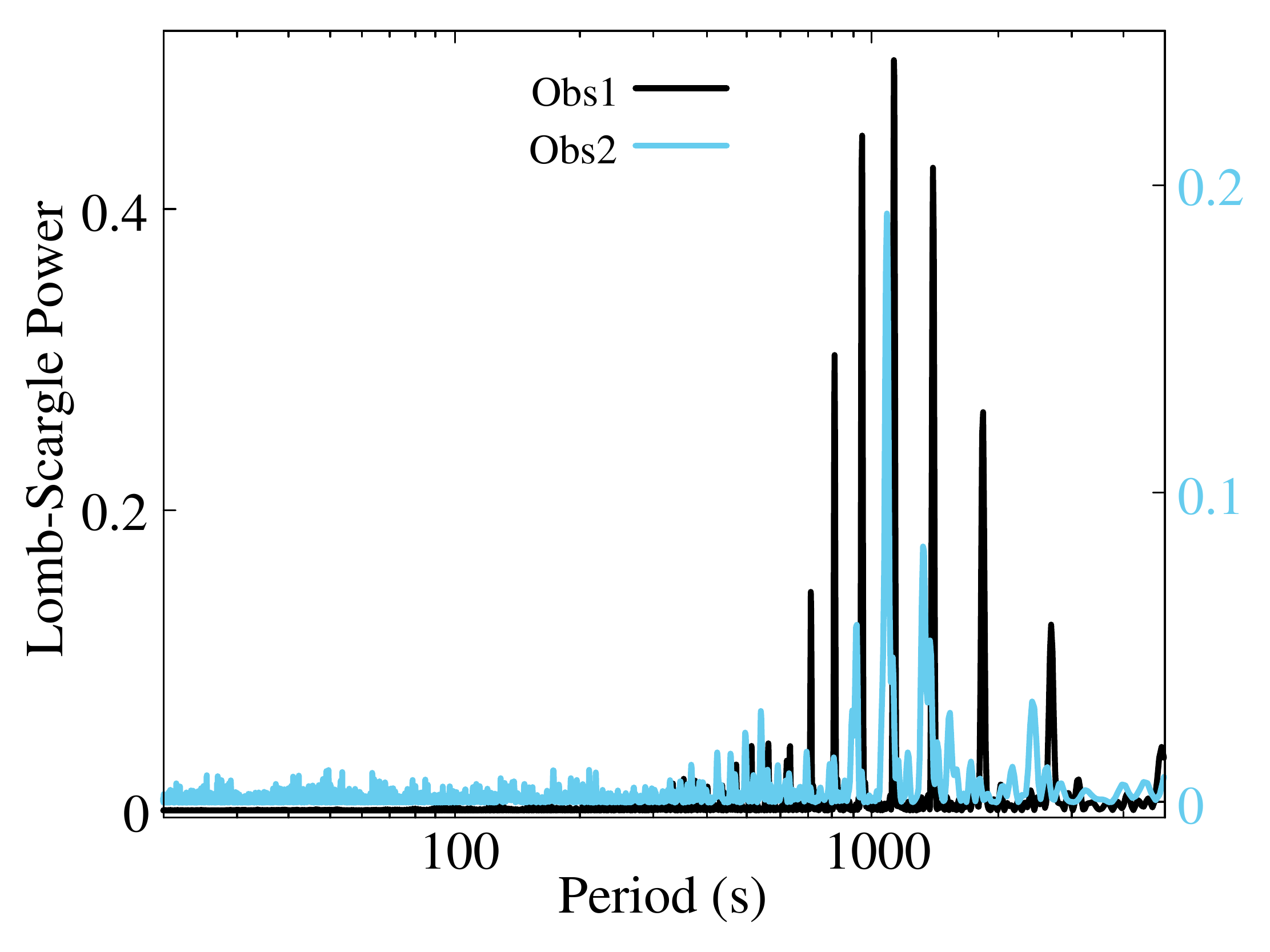}
\end{minipage}
\hspace{0.2\linewidth}
\begin{minipage}{0.30\textwidth}
\includegraphics[height=1.1\columnwidth]{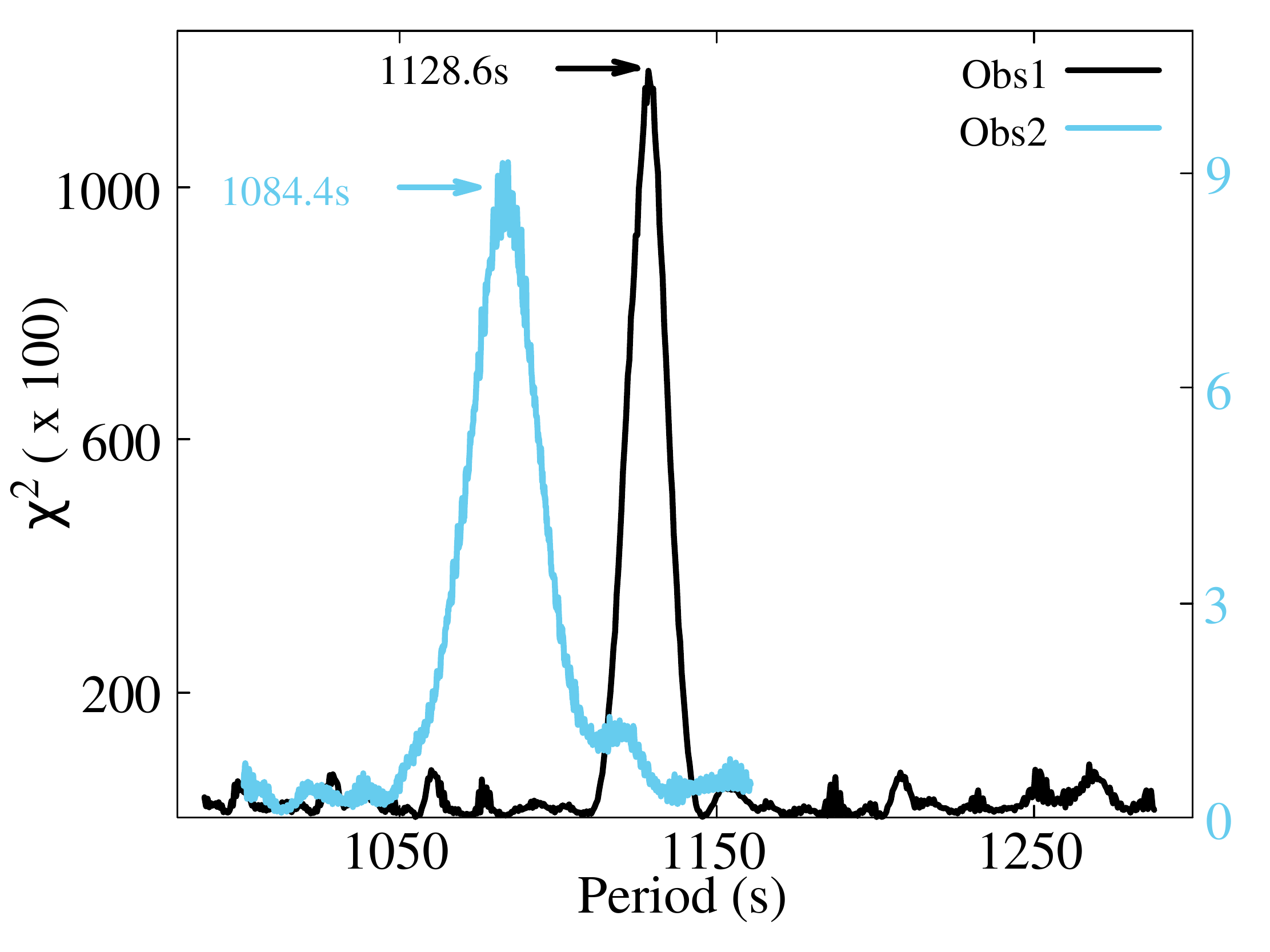}
\end{minipage}
\caption{\textit{Left} - Lomb-Scargle periodogram of the MAXI J0655--013 obtained using \emph{NuSTAR} data in 3-79 keV energy range. The peaks in the periodogram indicates the detection of pulsations, the peaks are observed at $\sim$ 1131 s and $\sim$ 1084 s for Obs1 (black) and Obs2 (cyan) respectively. \textit{Right} - Variation of the $\chi^{2}$ with pulse period obtained using \textsc{efsearch}. The period corresponding to the maximum $\chi^{2}$ corresponds to the best pulse period of the source.}
\label{fig3}
\end{figure*}

 We explored the periodicity in the light curves of the source using a Lomb-Scargle (LS) method \citep{1976Ap&SS..39..447L,1982ApJ...263..835S}. The LS periodograms of the pulsar are given on the left of Figure \ref{fig3}. For Obs1 we have used the cleaned event file obtained by CHU13 for the timing analysis, which has an exposure time 12.1 ks. Although CHU2 has an exposure time 17.1 ks but pulse profile obtained using it is not smooth. So we have not considered data from CHU2 for the timing analysis. Two peaks at $\sim$ 1131 s and $\sim$ 1084 s are observed in LS periodograms for Obs1 and Obs2 respectively. The pulse period obtained through LS method is further refined with the help of the epoch folding technique \citep{1987A&A...180..275L} \textsc{efsearch}. In the epoch folding technique, a $\chi^{2}$ value is computed for a given trial period ($P_{trial}$). Thereafter we plot the variation of $\chi^{2}$ with respect to the trial periods. In presence of pulsation a peak appears in the $\chi^{2}-P_{trial}$ plot. The trial period corresponding to this peak is the actual pulse period of the pulsar (as evident in Figure \ref{fig3}). The best period corresponding to the maximum $\chi^{2}$ are 1128.5$\pm$0.5 s and 1084.4$\pm$1.1 s for Obs1 and Obs2 respectively. The error attached with the pulse period was estimated using the method briefly described in \cite{Lut2012}. We simulated 500 light curves and for each simulated light curve, the pulse period is estimated using the epoch folding method. The standard deviation of the resulting best pulse period distribution yields the error associated with the pulse period. The pulse period estimated using the epoch folding method was used for subsequent analysis.

\subsection{Pulse profile}

\begin{figure*}
\begin{minipage}{0.32\textwidth}
\includegraphics[height=1.1\columnwidth]{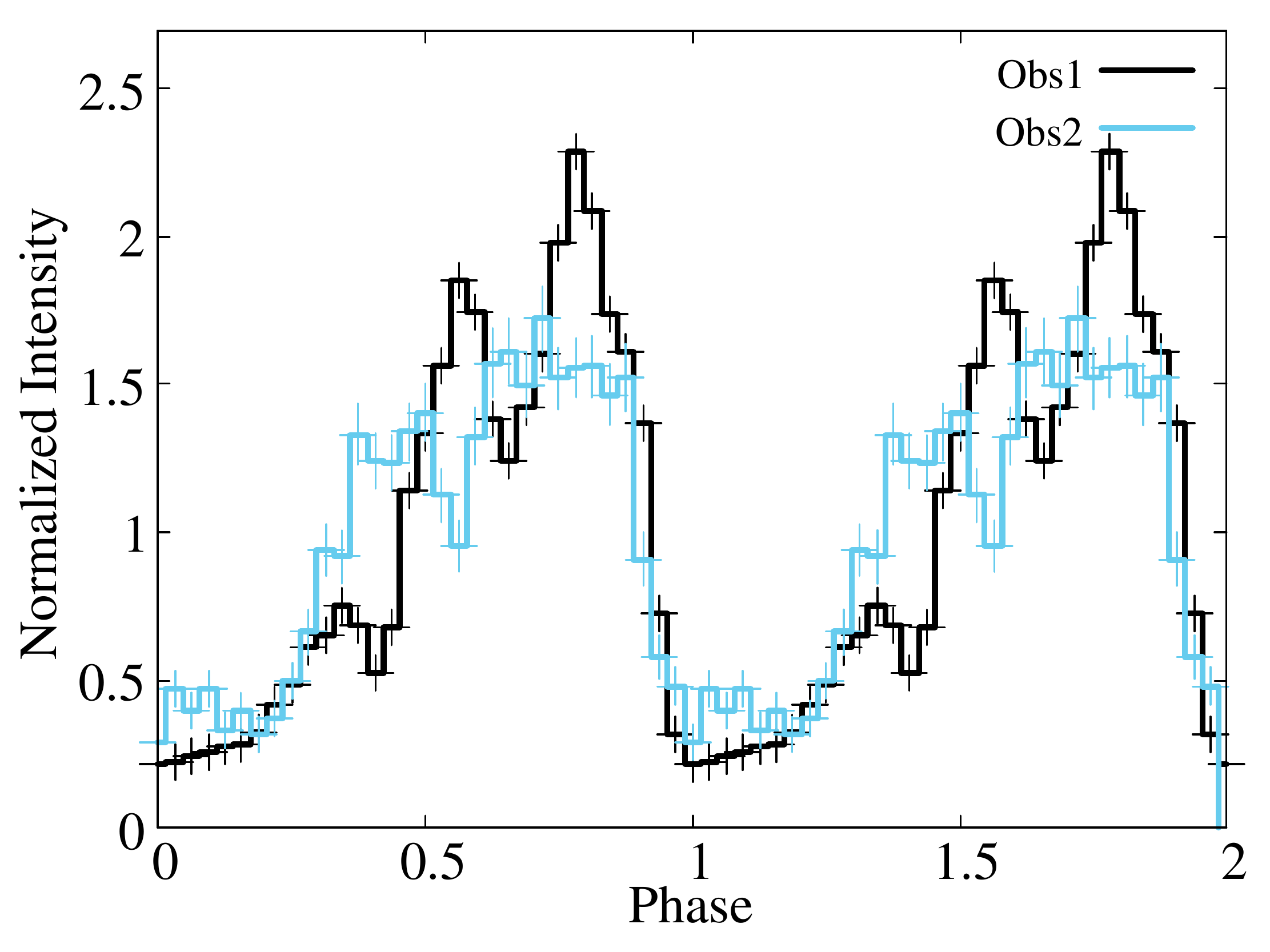}
\end{minipage}
\hspace{0.15\linewidth}
\begin{minipage}{0.34\textwidth}
\includegraphics[height=1.1\columnwidth]{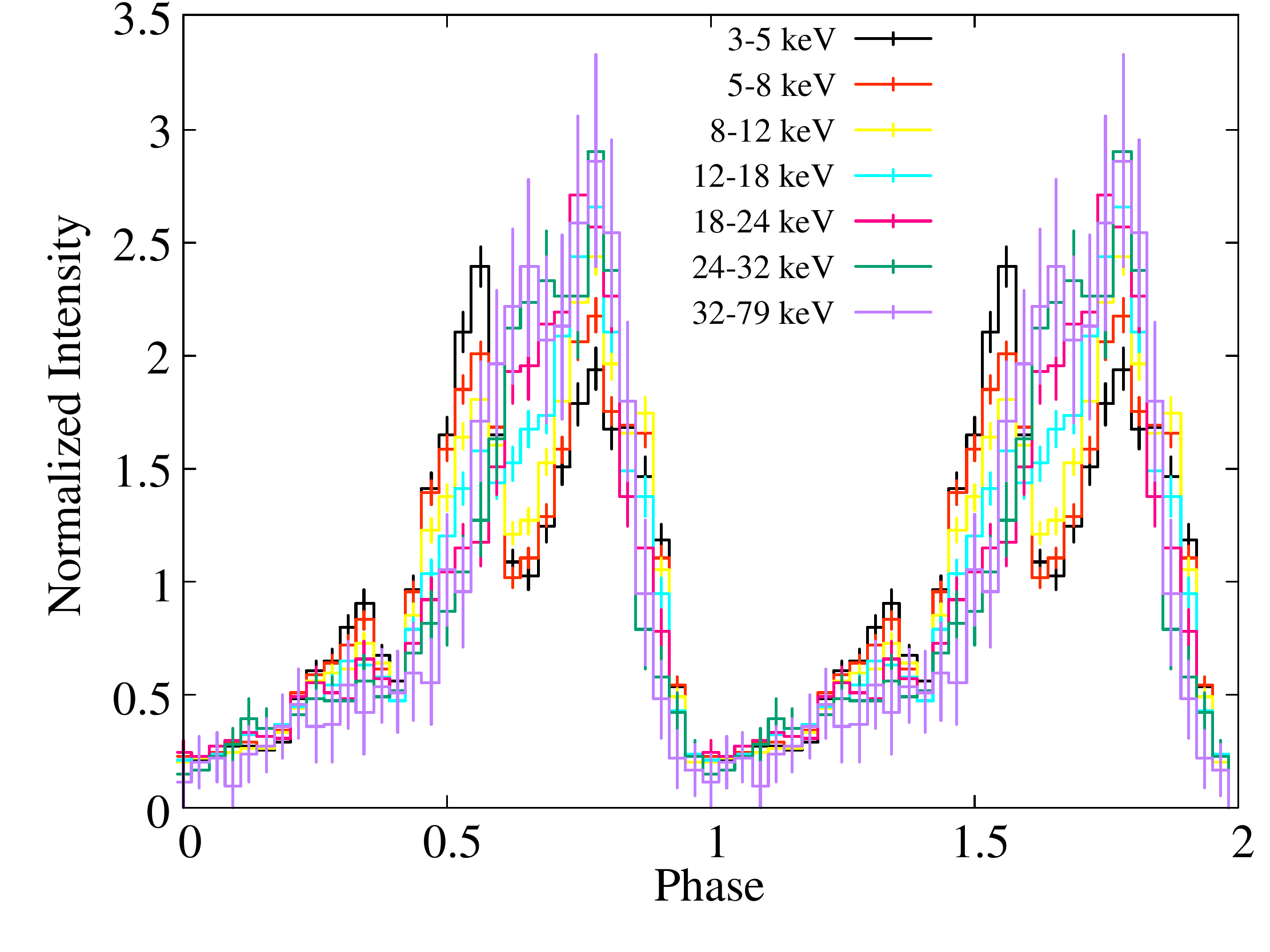}
\end{minipage}
\caption{\textit{Left} - Pulse profiles of the pulsar for Obs1 (black) and Obs2 (cyan) in 3-79 keV energy range. \textit{Right} - Pulse profiles of the pulsar in different energy ranges for Obs1.}
\label{fig4}
\end{figure*}

The pulse profile of the pulsar in the 3-79 keV energy range during Obs1 is two peaked (left of Figure \ref{fig4}). In the case of Obs2, a weak modulation of the intensity with phase is observed due to low count rates. To study the dependence of the pulse profile on energy, we folded the light curve of Obs1 of the source in different energy ranges about its pulse period. The pulse profiles for different energy bands are plotted on the right of Figure \ref{fig4}. The pulse profile of the pulsar shows some dependence on energy. For energy below 12 keV, the pulse profiles are double peaked and above it the pulse profiles become single peaked. From Figure \ref{fig4} one can observe that the height of the second peak increases, after 18 keV the first peak disappers and only this second peak remains in the pulse profile. The notch present at a pulse phase $\sim$0.4 is absent above 18 keV.

 The pulsed fraction, which is the relative amplitude of a pulse profile is defined as 
 \begin{equation}
PF_{max/min}=\dfrac{(F_{max}-F_{min})}{(F_{max}+F_{min})}
\label{eq1}
\end{equation}
where $F_{max}$ and $F_{min}$ are the pulse profile's maximum and minimum intensities, respectively. But the pulsed fraction obtained using the above method will add any noise present in the light curve to the actual pulsed fraction. For this, we take another form for the pulsed fraction given by

\begin{equation}
 PF_{rms}=\dfrac{\left(\dfrac{1}{N}\Sigma^{N}_{i=1}(p_{i}-\bar{p})^{2}\right)^{\dfrac{1}{2}}}{\bar{p}}
 \label{eq2}
\end{equation}

\begin{figure}
\includegraphics[scale=0.32]{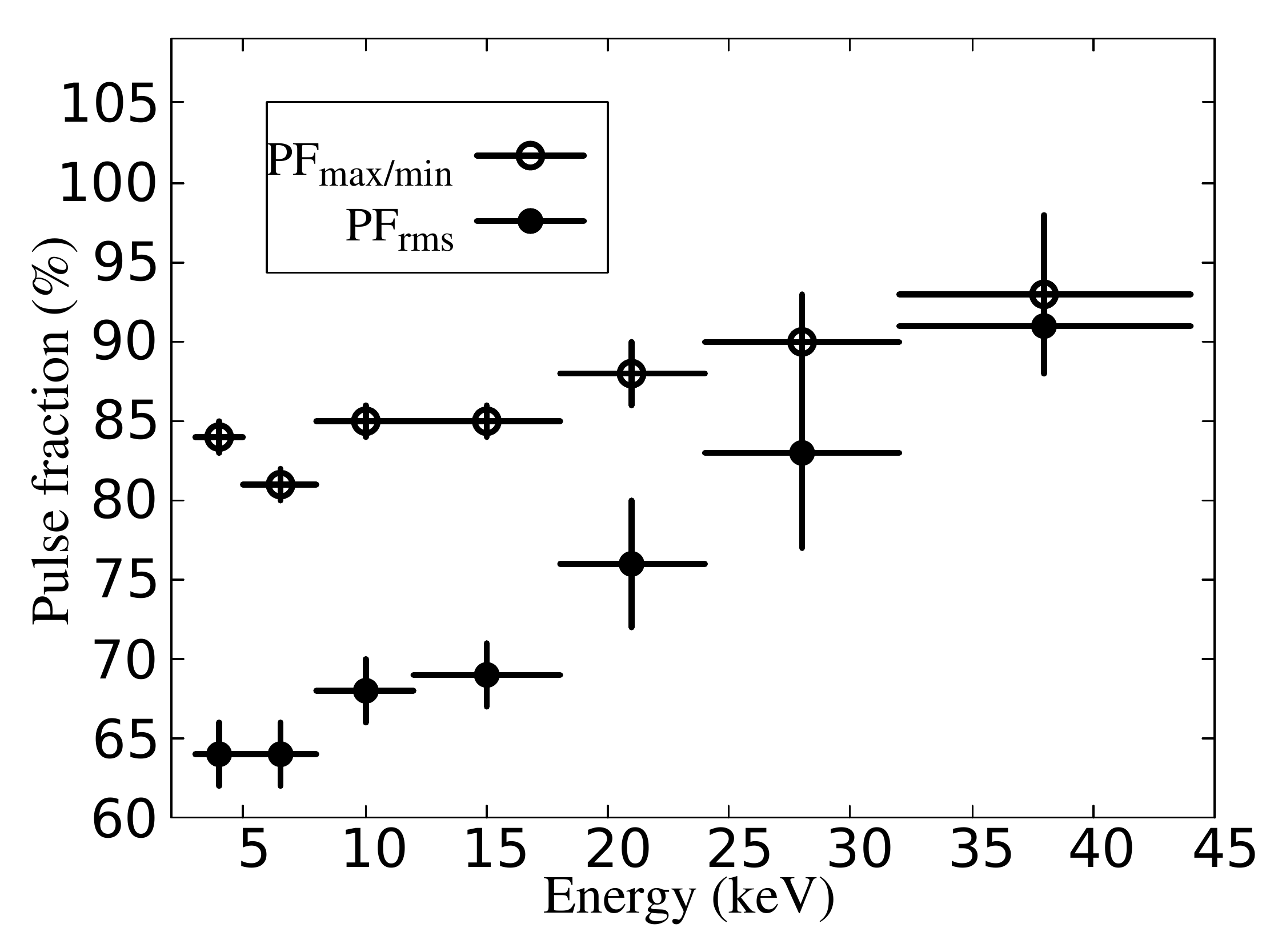}
\caption{Variation of the pulsed fraction of the pulsar with energy in the case of Obs1. The empty and filled circle represents $P_{max/min}$ and $P_{rms}$.}
\label{fig5}
\end{figure}

 The pulsed fractions calculated through this formula are relatively smaller than those calculated using the earlier definition (Equation \ref{eq1}). The variation of the pulsed fraction with energy is shown in Figure \ref{fig5}. The pulsed fraction of the pulsar increases with energy. The pulse profile of the pulsar becomes less structured as the energy increases, due to which the pulse fraction is found to increase in the hard energy range \citep{1997ApJS..113..367B}. \cite{2008AIPC.1054..191L} explained the increase in pulse fraction with the increase in energy considering a toy model. According to the model, the X-ray emitting region becomes more compact with an increase in energy, resulting in more pulsed emissions that emanate from the neutron star.

\subsection{Pulse Period evolution and Orbital Parameters}
\begin{figure}
\includegraphics[scale=0.38]{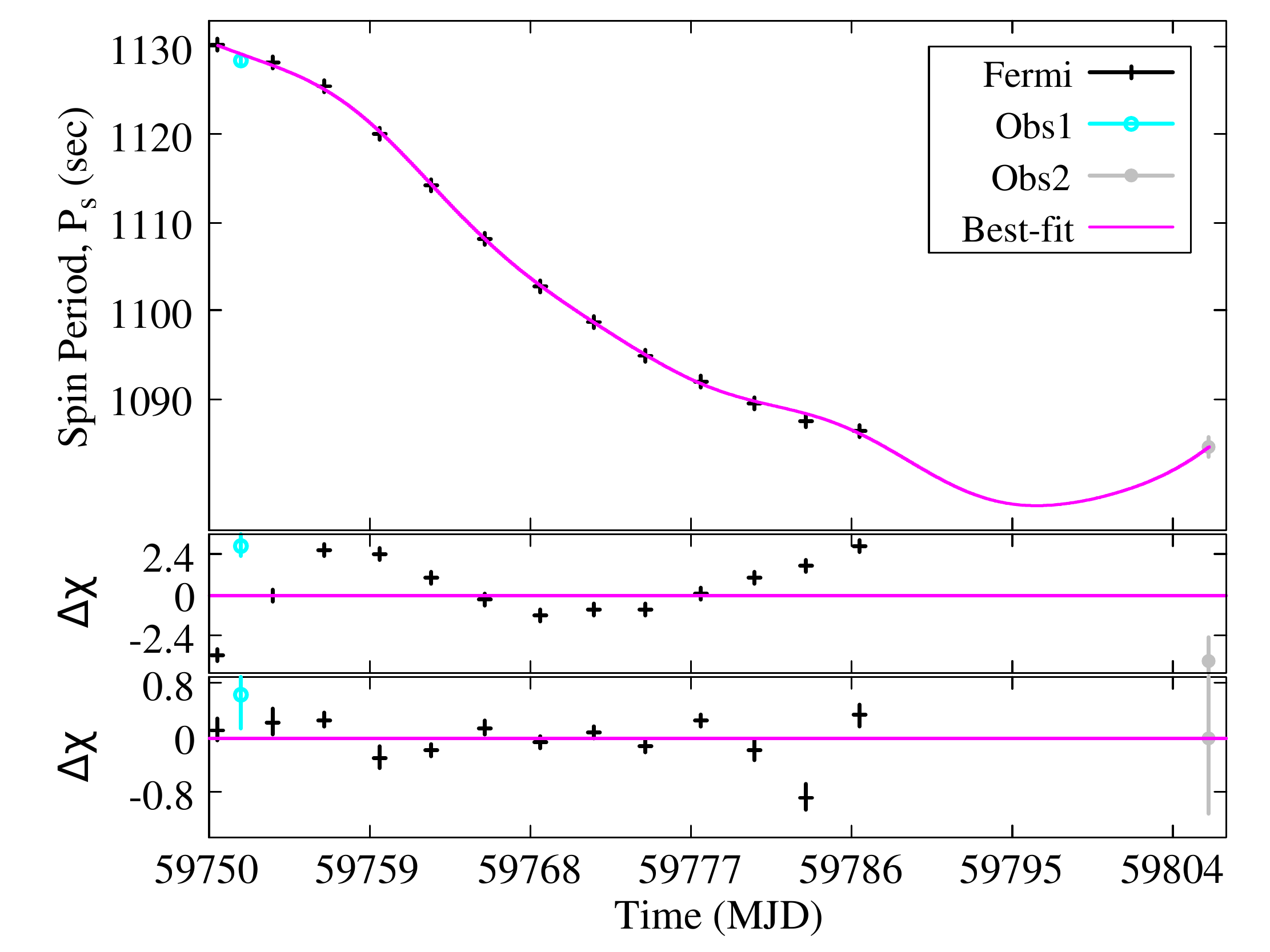}
\caption{The variation of the pulse period of MAXI J0655-013 with time is shown in the upper panel. The black, cyan, and grey points represent the spin period obtained from \emph{Fermi}, \emph{NuSTAR} Obs1 and Obs2 respectively. The magenta line is the best-fit line. The middle and upper panels in the figure represent the residuals (data-model) of the fitting without and with a model that accounts for the orbital modulation of the pulse period.}
\label{fig6}
\end{figure}

The pulse period of X-ray pulsars is found to vary throughout the outburst phase. Such variation in the pulse period during an outburst is also observed in MAXI J0655-013. The evolution of the pulse period of the pulsar during outburst was obtained by \emph{Fermi}-GBM accreting pulsar histories \footnote{\url{https://gammaray.nsstc.nasa.gov/gbm/science/pulsars.html}} and is shown in Figure \ref{fig6}. \emph{Fermi}-GBM continuously monitors accreting X-ray pulsars and keeps record of its pulse period history. \emph{Fermi}-GBM continuously monitors accreting X-ray pulsars and keeps records of their pulse period histories. A detailed study on the pulse period histories of persistent and transient X-ray pulsars observed by \emph{Fermi}-GBM is done by \cite{2020ApJ...896...90M}. The pulse period of the X-ray pulsar also varies with the orbital period due to the Doppler effect, and the pulsar is observed to spin faster when it approaches the observer and slowly when it recedes away from the observer. In the absence of orbital modulation, the pulse period of the pulsar evolves with time ($t$) as,
 
\begin{equation}
P(t)=P(t_{0})+\dot{P}(t-t_{0})+\ddot{P}(t-t_{0})^{2}
\label{eq3}
\end{equation}
where $P(t_{0})$ is the pulse period at the start time of the outburst ($t_{0}$), $\dot{P}$ and $\ddot{P}$ are the first and second derivatives of the pulse period. The observed time evolution of the spin period of an X-ray pulsar is modelled by the theoretical spin evolution given in \ref{eq3} along with the line of sight delay due to the orbital modulation from \cite{1981ApJ...247.1003D} to get the orbital parameters of the binary system \citep{2016MNRAS.457..258T}. We have rescaled the spin error in the case of GBM by a factor of 3 to balance the weights between GBM and \emph{NuSTAR} during the fitting process \citep{2017ApJ...843...69W}. However, to improve the curve fitting, we have included a cubic term $\dddot{P}(t-t_{0})^{3}$ in the Equation \ref{eq3}. The orbital parameters obtained from the fitting are presented in Table \ref{Tab2}. To decrease the uncertainty error associated with the time of periastron passage ($E$), we fix the two parameters $P(t_{0})$ and $\dddot{P}$ at the end of the fit.

\begin{table}
\begin{tabular}{c c}
\hline
Parameters & Values \\
\hline
$P(t_{0})$      & 1131.96 s  (fixed)     \\
$\dot{P}$    & -1.23    $\pm$ 0.02 s d$^{-1}$    \\
$\ddot{P}$& -0.0198    $\pm$ 0.0007 s d$^{-2}$  \\
$\dddot{P}$& 0.0005 s d$^{-3}$ (fixed) \\
Projected semi-major axis ($asini$)               &2.2          $\pm$ 0.2 lt-s\\    
$P_{orb}$             & 27.9          $\pm$ 1.7 d     \\
Time of periastron passage ($E$)               & 59750.1          $\pm$ 6.9  MJD   \\
Eccentricity ($e$)                &0.48        $\pm$ 0.30\\    
Angle of periastron ($w$)                &111.1         $\pm$ 14.1$^{\circ}$\\    
$\chi^{2}_{\nu}$/dof &            1.06/8     \\
\hline
\end{tabular}
\caption{Orbital parameters of MAXI J0655-013 obtained by fitting observed time evolution of the pulse period with the intrinsic spin evolution and orbital model. $\chi^{2}_{\nu}$ is the reduced $\chi^{2}$ of the fitting.}
\label{Tab2}
\end{table}

\subsection{Phase Averaged Spectral analysis}
\begin{figure*}
\begin{minipage}{0.30\textwidth}
\includegraphics[height=8cm, width=7 cm,angle=-90]{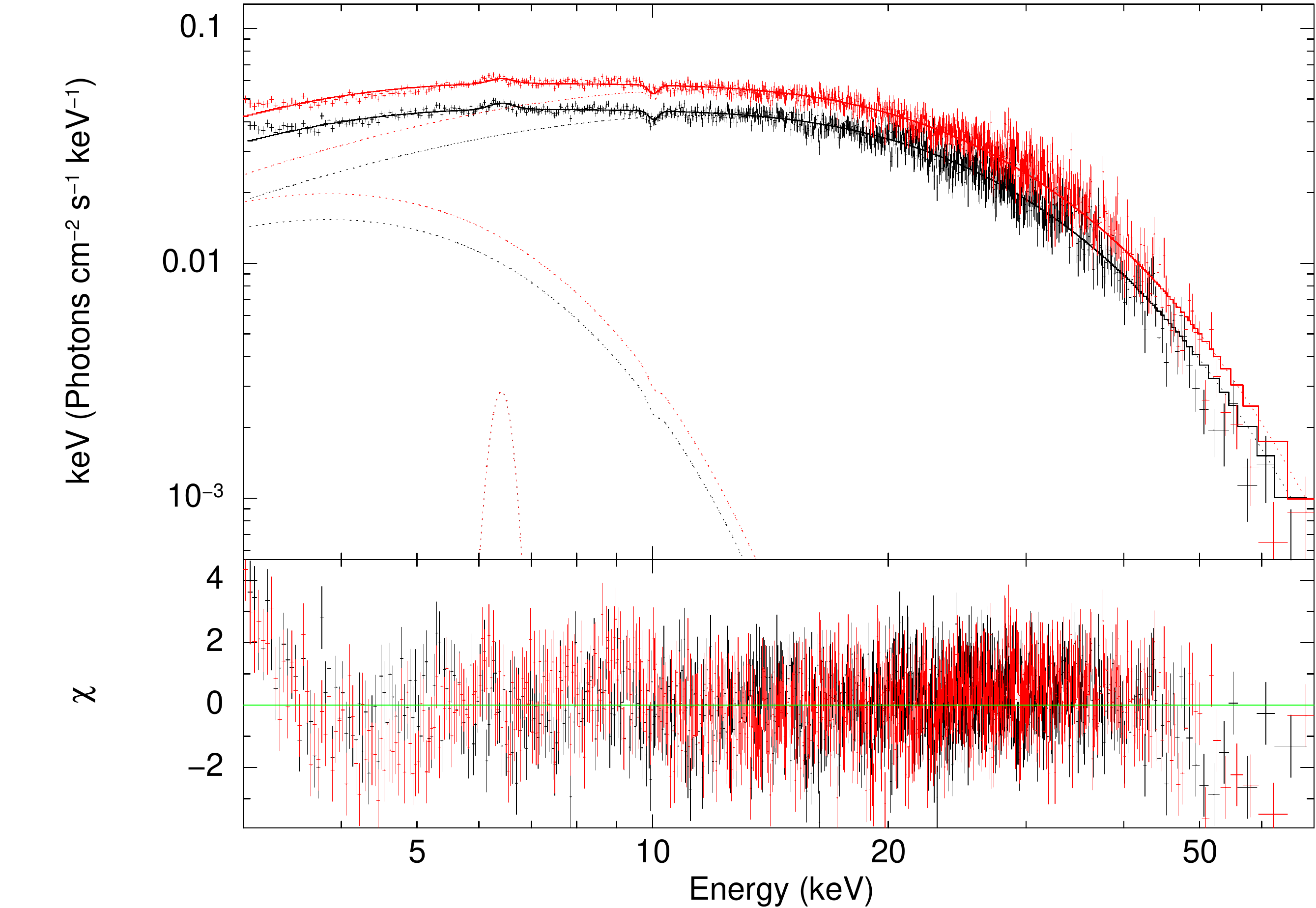}
\end{minipage}
\hspace{0.15\linewidth}
\begin{minipage}{0.30\textwidth}
\includegraphics[height=8cm, width=7 cm,angle=-90]{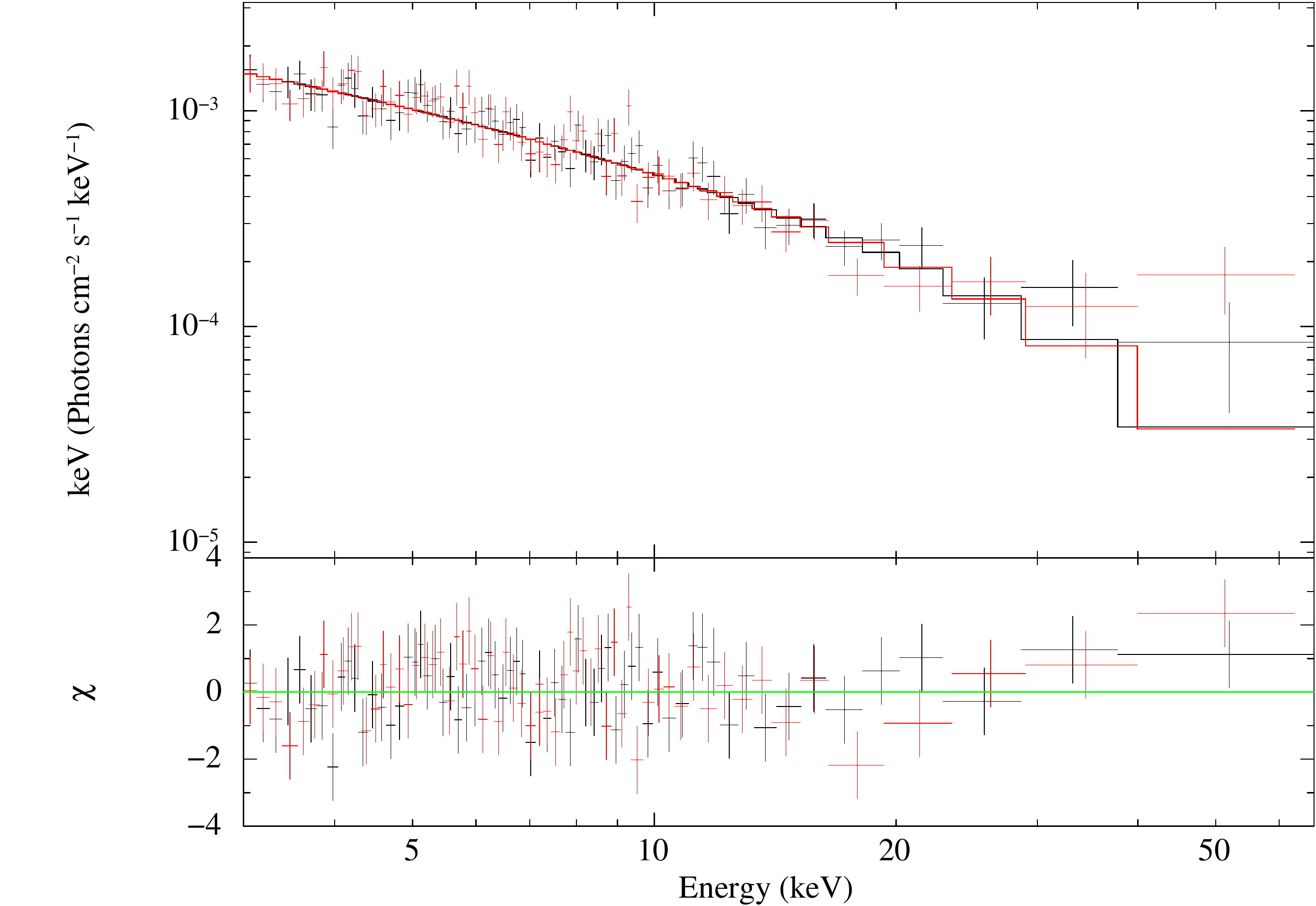}
\end{minipage}
\caption{\textit{Left}- The unfolded spectra of the pulsar for Obs1 is shown (upper panel). \textit{Right}- The unfolded spectra of the pulsar for Obs2. The red and black points are for FPMA and FPMB respectively. The lower panel shows the residuals of the fitting. The continuum model used for fitting is \textsc{cutoffpl}.}
\label{fig7}
\end{figure*}

The \emph{NuSTAR} FPMA and FPMB spectra for both observations Obs1 and Obs2 are grouped so that a minimum of 30 counts per bin can be taken. The spectra for the two \emph{NuSTAR} observations are fitted with a continuum model \textsc{cutoffpl}. In the case of Obs1, we used \textsc{blackbody} and \textsc{gaussian} models for analysis. The energy of the iron emission line is fixed at 6.4 keV, while its width was kept free to vary. The equivalent width of the iron emission line is found at $\sim$ 220 eV. An absorption-like feature is observed at $\sim$ 10 keV in Obs1, possibly due to the presence of Tungsten L-edge of the \emph{NuSTAR} detector \citep{2015ApJS..220....8M}. The absorption feature is modelled using a Gaussian absorption model \textsc{gabs}. The value of the column density along the direction of the source has been fixed to its expected magnitude 5.6$\times$10$^{21}$ cm$^{-2}$ \citep{2016A&A...594A.116H}, as the value of column density is not well constrained by \emph{NuSTAR} due to the absence of well-calibrated spectra below 3 keV. A \textsc{constant} model is introduced to account for the instrumental uncertainties between FPMA and FPMB. The \textsc{constant} model parameter corresponding to FPMA ($C_{FPMA}$) is set at 1, and for FPMB ($C_{FPMB}$) it was kept free to vary. The best-fit unfolded spectra along with the residuals are shown in Figure \ref{fig7}. In Obs2, there was no direct detection of the iron emission line in the spectrum of the source. However, we found that the upper limit of the equivalent width of the undetected iron line is 100 eV within the 90\% confidence interval (keeping the line energy of the iron emission line at 6.4 keV). The best-fit spectral parameters for both \emph{NuSTAR} observations are given in Table \ref{Tab3}. The flux estimates in the 3-79 keV energy range for Obs1 and Obs2 are $\sim$2.19$\times$10$^{-9}$ \unitfl and $\sim$1.38$\times$10$^{-11}$ \unitfl respectively. Assuming a distance of $\sim$4.1 kpc \citep{2022A&A...661A..38P}, we determine the luminosity in the given energy range to be $\sim$4.2$\times$10$^{36}$ \unilum and $\sim$3.9$\times$10$^{34}$ \unilum for Obs1 and Obs2 respectively.

\begin{table}
\scalebox{0.9}{
\begin{tabular}{c c c c}
\hline													
Model	&	Parameters	&		Obs1	&			Obs2			\\
\hline													
\textsc{constant}	&	$C_{FPMA}$	&			1	(fixed)		&	1	(fixed)	\\	
	&	$C_{FPMB}$	&			1.288	$\pm$	0.003	&	1.01	$\pm$	0.03	\\
\textsc{tbabs}	&	$nH$	($10^{21}$	cm$^{-2}$)	&	0.56	(fixed)		&	0.56	(fixed)	\\	
\textsc{bbody}	&	$kT$			&	1.32	$\pm$	0.02	&	\dots			\\
	&	norm			&	(1.8	$\pm$	0.1)$\times$10$^{-4}$	&	\dots			\\
\textsc{cutoffpl}	&	photon-index	($\alpha$)	&		-0.29	$\pm$	0.04	&	1.7	$\pm$	0.1	\\
	&	$E_{cut}$	(keV)	&	8.9	$\pm$	0.1	&	25	$\pm$8	\\	
	&	norm	&			(6.6	$\pm$	0.5)$\times$10$^{-3}$	&	(4$\pm$0.5)$\times$10$^{-3}$	\\
\textsc{gaussian}	&	$E_{Fe}$ (keV)			&	6.4	(fixed)		&	\dots			\\
	&	$\sigma_{Fe}$ (keV)			&	0.22	$\pm$	0.07	&	\dots			\\
	&	$norm$			&	(2.4	$\pm$	0.5)$\times$10$^{-4}$	&	\dots			\\
\textsc{gabs}& $E_{gabs}$ (keV)    &    10.03$\pm$0.05      &  \dots  \\
             &  $\sigma_{gabs}$ (keV) &  0.14$\pm$0.07      & \dots   \\
             &  $D_{gabs}$ &      0.03$\pm$0.01            & \dots    \\  
\hline													
	&	flux	(\unitfl)	&		(2.19	$\pm$	0.01)$\times$10$^{-9}$	&	(2.0	$\pm$	0.2)$\times$10$^{-11}$	\\
	&	$\chi^{2}$/dof	&		1876.1	/	1622	&	114	/125	\\	
\hline													

\hline													

\end{tabular}}
\caption{The best-fit parameters obtained from the spectral fitting. The flux is estimated in the 3-79 keV energy range. The norm has a unit of photons cm$^{-2}$ s$^{-1}$ keV$^{-1}$. }
\label{Tab3}
\end{table}

\subsection{Phase-Resolved Spectroscopy}

\begin{figure}
\includegraphics[scale=0.35]{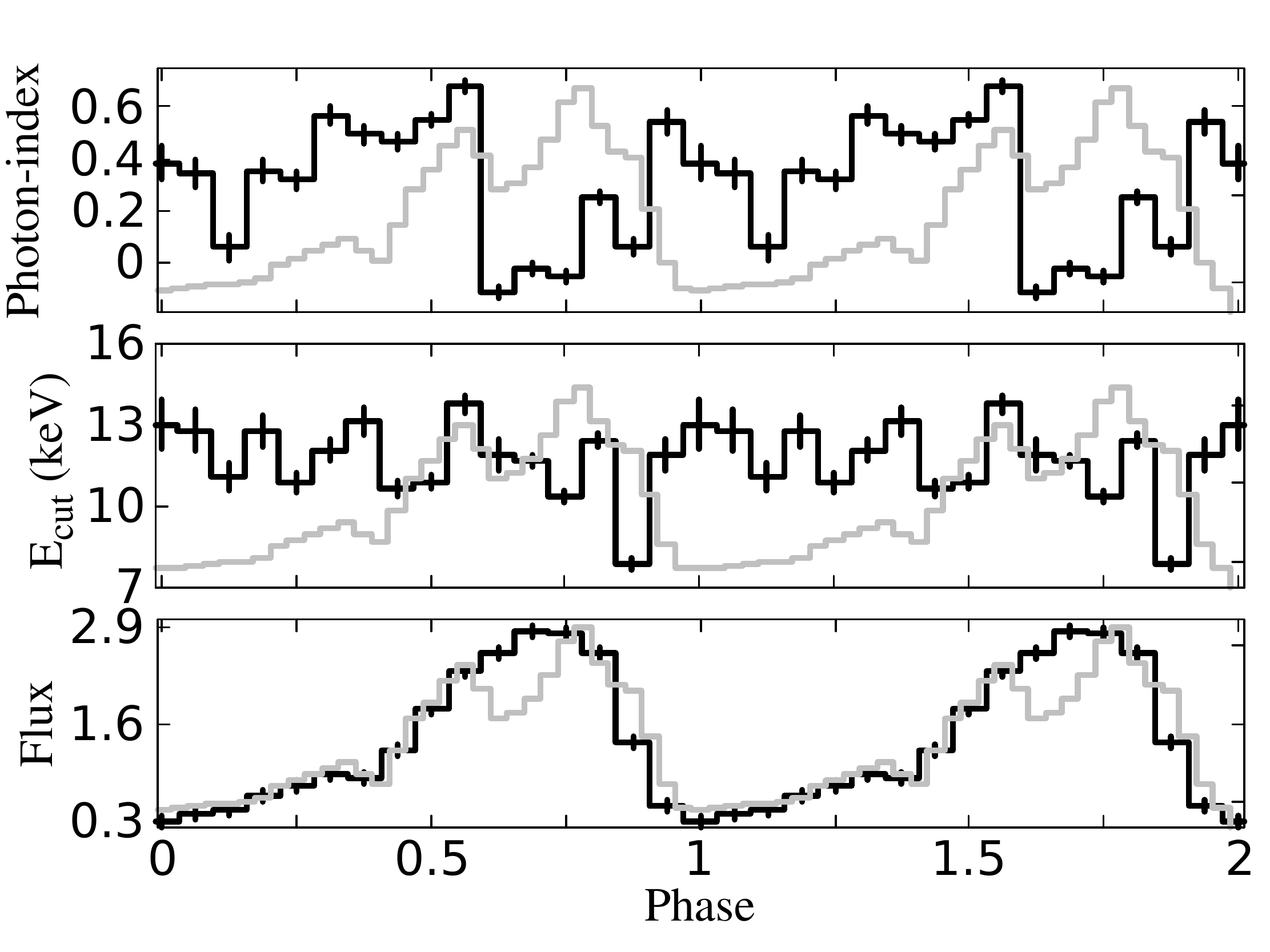}
\caption{Variation of photon-index (\textit{upper panel}), cutoff energy (\textit{middle panel}) and flux (\textit{lower panel}) with pulse phase. The flux is estimated in the 3-79 keV energy range and is in the order of 10$^{-10}$ erg cm$^{-2}$ s$^{-1}$. The figure in grey in all three panels is the pulse profile of the pulsar scaled suitably for plotting purposes.}
\label{fig8}
\end{figure}

For Obs1 we have extracted spectra for 16 independent phase bins with the help of estimated pulse period for all the CHU combinations mentioned in Table \ref{Tab1}. The spectra and response files from different CHU combinations are combined using the method described in Section 2. Due to a decrease in the signal-to-noise ratio of the phase-resolved spectra, no clear iron emission line was observed in the spectrum. Hence we fit the phase-resolved spectra using only the continuum model (\textsc{cutoffpl}). The value of the column density $(n_{H})$ is fixed at the expected value  $5.6\times$10$^{21}$ cm$^{-2}$, while fitting the spectrum. The photon index of the \textsc{cutoffpl} model is found to vary significantly with the pulse phase and lies between $\sim$ -0.1-0.6. The minimum value of the photon index is found to lie at the minimum between the two peaks of the pulse profile. We do not find systematic variation of the cutoff energy of the \textsc{cutoffpl}  with the pulse phase Figure \ref{fig8}.

\section{Discussion}

In the paper, we have studied the spectral and timing analysis of the recently discovered Be/X-ray pulsar MAXI J0655-013. It is a slowly rotating X-ray pulsar with a pulse period $>$ 1000 s. There are a few other X-ray pulsars with pulse periods greater than 1000 s, some of them are SXP1062 \citep{2012MNRAS.420L..13H}, SXP1323 \citep{2000A&AS..147...75S}, SXP4693 \citep{2010ApJ...716.1217L}, 2S 0114+650 \citep{1999ApJ...513L..45L}, and 4U 2206+54 \citep{Reig2012}. The iron emission line produced as a result of reprocessing of X-rays in the cold matter present in the vicinity of the neutron star is a common feature of BeXBs \citep{2013A&A...551A...1R}. The existence of an iron emission line gives direct evidence in favour of the presence of matter around the NS. We observed an iron emission line in the \emph{NuSTAR} spectra in the case of Obs1, however, no direct detection of the line is found in Obs2. We found the upper limit of the equivalent width of the 6.4 keV iron emission line in Obs2 is $\sim$ 100 eV (90\% confidence interval), which is consistent with the estimated equivalent width of the iron line in Obs1. The photon index of the cutoffpl model in Obs1 is less than that in Obs2, so it is clear that the emphNuSTAR spectrum of the pulsar is harder in the case of Obs1 than Obs2, as evident from the folded spectrum in Figure \ref{fig7}. The spectrum in the case of Obs2 is found to be steeper than Obs1. Generally, the spectrum of Be/X-ray pulsars varies with luminosity, and some of them show peculiar variations, like in GX 304-1 \citep{2019MNRAS.483L.144T}. In GX 304-1, the cutoff powerlaw spectrum at high luminosity state ($L_{X}\sim10^{36}-10^{37}$ \unilum) changes into a two-component spectrum at low luminosity state ($L_{X}\sim10^{34}$ \unilum). In the case of MAXI J0655-013, as the outburst slowly fades away, the thermally emitted X-rays become major contributors to the spectrum of the pulsar, as a result of which its spectrum becomes steep. Even if one models the Obs1 spectrum by a simple \textsc{cutoffpl} model \citep{2022ATel15495....1S} the photon index is about 0.3, which is significantly lower than its value estimated by us in Obs2. However, there are significant differences in the values of the spectral parameters computed by us and \citep{2022ATel15495....1S} for Obs1. The difference could be due to the difference in the model combinations used by \cite{2022ATel15495....1S} and us. \cite{2022ATel15495....1S} modelled the continuum spectrum only by \textsc{cutoffpl} model but we found that it is necessary to add \textsc{blackbody} and \textsc{gaussian} models to fit the Obs1 spectrum. 
The first \emph{NuSTAR} observation was done in 59751.84 MJD and during this observation, a pulse period of 1128.5$\pm$0.5 s was observed, and during the second observation which was done in 59805.94 MJD a pulse period of 1084.4$\pm$1.6 s was observed in the pulsar. Using the \emph{Fermi} \& \emph{NuSTAR} observations we found that the spin-up rate of the pulsar during an outburst is very high, $P\sim -1.23$ s d$^{-1}$. Such a high spin-up rate during an outburst is not observed in any other Be X-ray pulsars. \cite{10.1111/j.1365-2966.2009.15631.x} studied the pulse period change in 24 X-ray pulsars of Small Magellanic Cloud (SMC) having pulse periods between 2.372-1352 s. Out of 24 pulsars they observed short-term (50-500 days) spin change in 15 pulsars (four showed spin-down). The maximum short-term spin-up rate is -0.024 s d$^{-1}$ observed in SXP892 \cite{10.1111/j.1365-2966.2009.15631.x}. The spin-up rate obtained here is almost 35 times higher than that observed in SXP892. The peak luminosity of the outburst in SXP892, when this short-term spin-up rate is observed, was 4.8$\times$10$^{36}$ \unilum \citep{10.1111/j.1365-2966.2009.15631.x}. Generally, it is observed that during an outburst the spin of the pulsar goes up due to the large spin-up torque \citep{Bildsten_1997}, but we have noted here that in the case of MAXI J0655--013 the spin-up rate is very high. \cite{2012A&A...537L...1H} reported a high short-term spin-down rate of 0.26 s d$^{-1}$ in SXP1062 and explained the phenomenon in terms of large spin-down accretion torque. The spin-up rate in MAXI J0655--013 is almost five times greater than the spin-down rate in the former case. However, the long-term spin-up/spin-down of pulsars is smaller than the short-term change because the pulsar may pass through intermittent spin-down/spin-up phases \citep{10.1111/j.1365-2966.2009.15631.x}. To estimate the luminosity at the peak of the outburst, we converted the \emph{MAXI}/GSC count rate in 2-20 keV into flux in the same energy range using the WebPIMMS system \footnote{\url{https://heasarc.gsfc.nasa.gov/cgi-bin/Tools/w3pimms/w3pimms.pl}}. We have used the spectral parameters for Obs1 given in Table \ref{Tab3} and column density was fixed to the galactic value of $\sim$5.6$\times$10$^{21}$ cm$^{-2}$ while converting count rate into flux. Using the bolometric flux of 2.3$\times$10$^{-9}$ \unitfl in 0.1-100 keV for Obs1, we obtained a bolometric correction factor ($K_{bol}$) for \emph{MAXI} flux in 2-20 keV, which is $\sim$1.2. The peak bolometric luminosity was found at $\sim$ 6.9$\times$10$^{37}$ \unilum. The peak luminosity of the outburst in MAXI J0655--013 is almost 100 and 10 times greater than the maximum luminosity when the large short-term spin change was observed in SXP1062 \citep{2012A&A...537L...1H} and SXP892 \citep{10.1111/j.1365-2966.2009.15631.x} respectively. The large spin-up rate in MAXI J0655--013 is probably due to the large spin-up torque that is acting on the NS. It is worth mentioning here that a long spin-up rate of about -29.9 s yr$^{-1}$ (-0.082 s d$^{-1}$ ) was observed in SXP1323 \citep{2022A&A...661A..33M}.  

The magnetic field of the pulsar can be estimated using the torque model of a standard disk accretion given by \cite{1979ApJ...232..259G,1979ApJ...234..296G}, if the spin-up rate is known. The spin-up rate ($\dot{P}$) obtained from the torque model is found to depend on the luminosity as,
\begin{equation}
\dot{P} = -5\times 10^{-5}k^{1/2}n(\omega_{s})M_{1.4}^{-3/7}R_{6}^{6/7}I_{45}^{-1}\mu_{30}^{2/7}(PL_{37}^{3/7})^2 \mbox{ s yr$^{-1}$ }
\label{Eq4}
\end{equation}
where $k$ is a parameter that accounts for the details of the interaction between the accretion flow and the magnetosphere, its typical value is about 0.5 \citep{1979ApJ...232..259G}, $M_{1.4}$ is the mass of the NS in terms of solar mass and $R_{6}$ is the radius (R) of the NS divided by 10$^{6}$ cm, $P$ is the pulse period in seconds, $I_{45}$ is the moment of inertia in the order of 10$^{45}$ g cm$^{2}$, $L_{37}$ is the luminosity in the order of 10$^{37}$ \unilum, $\mu_{30}$ is the magnetic moment in the unit of 10$^{30}$ G cm$^{3}$, $n(\omega_{s})$ is the dimensionless torque, and $\omega_{s}$ is the fastness parameter, which is defined as,

\begin{equation}\omega_{s} = 3.3k^{3/2}M_{1.4}^{-2/7}R_{6}^{-3/7}\mu_{30}^{6/7} (PL_{37}^{3/7})^{-1}\end{equation}When $\omega_{s}$ is such that 0<$\omega_{s}$<0.9 then we approximate the dimensionless torque and is given by,
\begin{equation}
n(\omega_{s})=1.4(1-2.86\omega_{s})(1-\omega_{s})^{-1}
\end{equation}

\cite{2014MNRAS.437.3664H} gave a good approximation of \ref{Eq4}, which is
\begin{equation}\dot{P} = -7.1\times 10^{-5}k^{1/2}M_{1.4}^{-3/7}R_{6}^{6/7}I_{45}^{-1}(1-\omega_{s})\mu_{30}^{2/7}(PL_{37}^{3/7})^2 \mbox{ s yr$^{-1}$ }
\label{Eq7}
\end{equation}

Using $M_{1.4}=1$, $R_{6}=1$, $I_{45}=1$, $k=0.5$, $P\sim1100$ s, peak luminosity $L_{37}=6.9$ and $\dot{P}\sim-1.23$\;s d$^{-1}$ in Equation \ref{Eq7} we get $\mu_{30}\sim4$. As $\mu\sim\dfrac{1}{2}BR^{3}$, the magnetic field of the pulsar is $\sim$8$\times$10$^{12}$ G. However this estimated magnetic field is affected by uncertainty on the distance, mass, and radius of the neutron star. It is known that for a pulsar with a spin period $P_{s} \gtrsim$ 1000 s, the estimated magnetic field is very high ($>$ 10$^{13}$ G) e.g. - SXP 1062 \citep{2012ApJ...757..171F}, SXP 1323 \citep{2022A&A...661A..33M}, 2S 0114+650 \citep{1999ApJ...513L..45L} and 4U 2206+54 \citep{Reig2012}. To explain the long spin period ($\sim$ 2.7 hr) in 2S 0144+650 \cite{1999ApJ...513L..45L} proposed that its magnetic field be greater than 10$^{14}$ G. \cite{Reig2012} also found that for a pulsar to have $P_{s} > 1000 s$, a high magnetic field like the one observed in the magnetar is required.It is possible that the X-ray pulsars like MAXI J0655--013 with $P>1000$ s and high magnetic field $B>10^{13}$ G may never reach a propeller phase after an outburst, as argued by \cite{2022A&A...661A..33M}, instead the X-ray source enters into a phase of stable accretion from a cold disk phase \citep{2017A&A...608A..17T}. In this accretion phase, the mass accretion rate becomes so small that the temperature of the disc becomes lower than the hydrogen recombination limit (T$\sim$6500 K). The rapid fading of the intensity of the source stops at this phase, and stable accretion may start from a cold disc. The viscosity in the cold disk is small, and the slow accretion is expected to continue \citep{2017A&A...608A..17T}. So it can be argued that during \emph{NuSTAR} Obs2, the X-ray pulsar might have entered a state of stable cold disk accretion. 
 The pulse profiles of the source are simple. The pulse profile shows a strong dependence on the energy and changes its shape from a double to a single peak as we move from the soft to the hard X-ray range. This change in the pulse profile with energy can be due to the disappearance of the dips at a phase $\sim$ 0.55 in between the two peaks with an increase in energy. The dips in certain phases, especially in the soft X-ray range, can be due to the absorption of the soft X-rays by narrow streams of matter phase-locked with the neutron star \citep{2016MNRAS.457.2749J}. The luminosity of the pulsar in the 3-79 keV energy range was about $\sim$4.2$\times$10$^{36}$ \unilum and $\sim$3.9$\times$10$^{34}$ \unilum for Obs1 and Obs2 respectively, so we cannot expect an extended accretion column to form above the NS surface \citep{2015MNRAS.447.1847M,10.1093/mnras/stv2087}. Then the X-ray photons are expected to originate from the surface of NS or close to it. The resulting emission pattern is a pencil beam \citep{1976MNRAS.175..395B}. At high luminosity, we expect an accretion column to form above the accretion column, and the corresponding emission pattern leads to the fan beam emission pattern. The two beam emission patterns are related to two different accretion states, separated by a luminosity known as critical luminosity $(L_{crit})$ \citep{1976MNRAS.175..395B}.  For luminosities above the critical luminosity (super-critical accretion), the infalling plasma is stopped above the surface of the neutron star by a radiation-dominated shock. The matter is slowly decelerated as it moves through the shock, and the X-ray photon escapes through the side wall of the accretion column, resulting in a fan-shaped beaming pattern \citep{2012A&A...544A.123B}. The fan beam emission pattern results in complex pulse profiles with multiple peaks. For luminosity below the critical luminosity (sub-critical accretion), the formation of the radiation-dominated shock wave does not occur, and the radiation emitted from the surface of the NS moves perpendicular to the surface, forming a pencil beam. The resulting pulse profiles will be simple. For a typical X-ray pulsar, the magnetic field is of the order of 10$^{12}$ G, and in such a case, $L_{crit}$ during the bright stage is of the order of 10$^{37}$ \unilum \citep{2012A&A...544A.123B}. However, for pulsars having a magnetic field of the order of 10$^{13}$ G the value of the critical luminosity is a magnitude higher than its value for typical X-ray pulsars. So in the case of MAXI J0655-013 with an estimated magnetic field $\sim$ 8$\times$10$^{12}$ G, the critical luminosity of the pulsar is expected to be greater than the luminosity of the pulsar during \emph{NuSTAR} observations. The simple pulse profiles of the source also hint to us that it is in the sub-critical regime of accretion.

\section*{Summary}
 We report the first characteristic study of recently discovered Be/X-ray MAXI J0655-013 using the two \emph{NuSTAR} observations. It is a slow rotating pulsar with a  short-term spin-up rate of -1.23 s d$^{-1}$. The pulse profile of the first \emph{NuSTAR} observation of the pulse shows dependence on energy. The pulse profile of the pulsar is simple indicating that the X-ray emission occurs from the neutron star's surface or close to it. Spectral evolution with time is observed from the two \emph{NuSTAR} observations.
 
 \section*{Data availability}
 
 The observational data of NuSTAR used in this study can be accessed from the HEASARC data archive and is publicly available for carrying out research work.

\section*{Acknowledgement}

This work made use of \emph{NuSTAR} data downloaded from NASA the high-energy astrophysics science archive research center (HEASARC) data archive. \emph{NuSTAR} is a project led by Caltech and funded by NASA. This research also has made use of the NuSTAR data analysis software (NUSTARDAS) jointly developed by the ASI science data center (ADSC), Italy, and Caltech. We have also made use of the pulse period history provided by \emph{Fermi} GBM team. This research has made use of the software provided by HEASARC, which is supported by the Astrophysics Division at NASA/GSFC and the high energy astrophysics division of the Smithsonian Astrophysical Observatory (SAO). We are thankful for the valuable suggestions and comments of the anonymous reviewer, which has helped us to improve the quality of the manuscript.



\bibliographystyle{mnras}
\bibliography{maxi}

\begin{thebibliography}{}
\makeatletter
\relax
\def\mn@urlcharsother{\let\do\@makeother \do\$\do\&\do\#\do\^\do\_\do\%\do\~}
\def\mn@doi{\begingroup\mn@urlcharsother \@ifnextchar [ {\mn@doi@}
  {\mn@doi@[]}}
\def\mn@doi@[#1]#2{\def\@tempa{#1}\ifx\@tempa\@empty \href
  {http://dx.doi.org/#2} {doi:#2}\else \href {http://dx.doi.org/#2} {#1}\fi
  \endgroup}
\def\mn@eprint#1#2{\mn@eprint@#1:#2::\@nil}
\def\mn@eprint@arXiv#1{\href {http://arxiv.org/abs/#1} {{\tt arXiv:#1}}}
\def\mn@eprint@dblp#1{\href {http://dblp.uni-trier.de/rec/bibtex/#1.xml}
  {dblp:#1}}
\def\mn@eprint@#1:#2:#3:#4\@nil{\def\@tempa {#1}\def\@tempb {#2}\def\@tempc
  {#3}\ifx \@tempc \@empty \let \@tempc \@tempb \let \@tempb \@tempa \fi \ifx
  \@tempb \@empty \def\@tempb {arXiv}\fi \@ifundefined
  {mn@eprint@\@tempb}{\@tempb:\@tempc}{\expandafter \expandafter \csname
  mn@eprint@\@tempb\endcsname \expandafter{\@tempc}}}

\bibitem[\protect\citeauthoryear{{Basko} \& {Sunyaev}}{{Basko} \&
  {Sunyaev}}{1976}]{1976MNRAS.175..395B}
{Basko} M.~M.,  {Sunyaev} R.~A.,  1976, \mn@doi [\mnras]
  {10.1093/mnras/175.2.395}, \href
  {https://ui.adsabs.harvard.edu/abs/1976MNRAS.175..395B} {175, 395}

\bibitem[\protect\citeauthoryear{{Becker} et~al.,}{{Becker}
  et~al.}{2012}]{2012A&A...544A.123B}
{Becker} P.~A.,  et~al., 2012, \mn@doi [\aap] {10.1051/0004-6361/201219065},
  \href {https://ui.adsabs.harvard.edu/abs/2012A&A...544A.123B} {544, A123}

\bibitem[\protect\citeauthoryear{{Bildsten} et~al.,}{{Bildsten}
  et~al.}{1997a}]{1997ApJS..113..367B}
{Bildsten} L.,  et~al., 1997a, \mn@doi [\apjs] {10.1086/313060}, \href
  {https://ui.adsabs.harvard.edu/abs/1997ApJS..113..367B} {113, 367}

\bibitem[\protect\citeauthoryear{Bildsten et~al.,}{Bildsten
  et~al.}{1997b}]{Bildsten_1997}
Bildsten L.,  et~al., 1997b, \mn@doi [The Astrophysical Journal Supplement
  Series] {10.1086/313060}, 113, 367

\bibitem[\protect\citeauthoryear{Coe, McBride  \& Corbet}{Coe
  et~al.}{2009}]{10.1111/j.1365-2966.2009.15631.x}
Coe M.~J.,  McBride V.~A.,   Corbet R. H.~D.,  2009, \mn@doi [Monthly Notices
  of the Royal Astronomical Society] {10.1111/j.1365-2966.2009.15631.x}, 401,
  252

\bibitem[\protect\citeauthoryear{{Deeter}, {Boynton}  \& {Pravdo}}{{Deeter}
  et~al.}{1981}]{1981ApJ...247.1003D}
{Deeter} J.~E.,  {Boynton} P.~E.,   {Pravdo} S.~H.,  1981, \mn@doi [\apj]
  {10.1086/159110}, \href
  {https://ui.adsabs.harvard.edu/abs/1981ApJ...247.1003D} {247, 1003}

\bibitem[\protect\citeauthoryear{{Fu} \& {Li}}{{Fu} \&
  {Li}}{2012}]{2012ApJ...757..171F}
{Fu} L.,  {Li} X.-D.,  2012, \mn@doi [\apj] {10.1088/0004-637X/757/2/171},
  \href {https://ui.adsabs.harvard.edu/abs/2012ApJ...757..171F} {757, 171}

\bibitem[\protect\citeauthoryear{{Ghosh} \& {Lamb}}{{Ghosh} \&
  {Lamb}}{1979a}]{1979ApJ...232..259G}
{Ghosh} P.,  {Lamb} F.~K.,  1979a, \mn@doi [\apj] {10.1086/157285}, \href
  {https://ui.adsabs.harvard.edu/abs/1979ApJ...232..259G} {232, 259}

\bibitem[\protect\citeauthoryear{{Ghosh} \& {Lamb}}{{Ghosh} \&
  {Lamb}}{1979b}]{1979ApJ...234..296G}
{Ghosh} P.,  {Lamb} F.~K.,  1979b, \mn@doi [\apj] {10.1086/157498}, \href
  {https://ui.adsabs.harvard.edu/abs/1979ApJ...234..296G} {234, 296}

\bibitem[\protect\citeauthoryear{{HI4PI Collaboration} et~al.,}{{HI4PI
  Collaboration} et~al.}{2016}]{2016A&A...594A.116H}
{HI4PI Collaboration} et~al., 2016, \mn@doi [\aap]
  {10.1051/0004-6361/201629178}, \href
  {https://ui.adsabs.harvard.edu/abs/2016A&A...594A.116H} {594, A116}

\bibitem[\protect\citeauthoryear{{Haberl}, {Sturm}, {Filipovi{\'c}}, {Pietsch}
  \& {Crawford}}{{Haberl} et~al.}{2012}]{2012A&A...537L...1H}
{Haberl} F.,  {Sturm} R.,  {Filipovi{\'c}} M.~D.,  {Pietsch} W.,   {Crawford}
  E.~J.,  2012, \mn@doi [\aap] {10.1051/0004-6361/201118369}, \href
  {https://ui.adsabs.harvard.edu/abs/2012A&A...537L...1H} {537, L1}

\bibitem[\protect\citeauthoryear{Harrison et~al.,}{Harrison
  et~al.}{2013}]{Harrison_2013}
Harrison F.~A.,  et~al., 2013, \mn@doi [The Astrophysical Journal]
  {10.1088/0004-637x/770/2/103}, 770, 103

\bibitem[\protect\citeauthoryear{{H{\'e}nault-Brunet}
  et~al.,}{{H{\'e}nault-Brunet} et~al.}{2012}]{2012MNRAS.420L..13H}
{H{\'e}nault-Brunet} V.,  et~al., 2012, \mn@doi [\mnras]
  {10.1111/j.1745-3933.2011.01183.x}, \href
  {https://ui.adsabs.harvard.edu/abs/2012MNRAS.420L..13H} {420, L13}

\bibitem[\protect\citeauthoryear{{Ho}, {Klus}, {Coe}  \& {Andersson}}{{Ho}
  et~al.}{2014}]{2014MNRAS.437.3664H}
{Ho} W. C.~G.,  {Klus} H.,  {Coe} M.~J.,   {Andersson} N.,  2014, \mn@doi
  [\mnras] {10.1093/mnras/stt2193}, \href
  {https://ui.adsabs.harvard.edu/abs/2014MNRAS.437.3664H} {437, 3664}

\bibitem[\protect\citeauthoryear{{Jaisawal}, {Naik}  \& {Epili}}{{Jaisawal}
  et~al.}{2016}]{2016MNRAS.457.2749J}
{Jaisawal} G.~K.,  {Naik} S.,   {Epili} P.,  2016, \mn@doi [\mnras]
  {10.1093/mnras/stw085}, \href
  {https://ui.adsabs.harvard.edu/abs/2016MNRAS.457.2749J} {457, 2749}

\bibitem[\protect\citeauthoryear{{Joye} \& {Mandel}}{{Joye} \&
  {Mandel}}{2003}]{2003ASPC..295..489J}
{Joye} W.~A.,  {Mandel} E.,  2003, in {Payne} H.~E.,  {Jedrzejewski} R.~I.,
  {Hook} R.~N.,  eds,  Astronomical Society of the Pacific Conference Series
  Vol. 295, Astronomical Data Analysis Software and Systems XII. p.~489

\bibitem[\protect\citeauthoryear{{Kennea}}{{Kennea}}{2022}]{2022ATel15564....1K}
{Kennea} J.~A.,  2022, The Astronomer's Telegram, \href
  {https://ui.adsabs.harvard.edu/abs/2022ATel15564....1K} {15564, 1}

\bibitem[\protect\citeauthoryear{{Kennea}, {Gronwall}, {Page}, {Palmer},
  {Tohuvavohu}  \& {Neil Gehrels Swift Observatory Team}}{{Kennea}
  et~al.}{2022a}]{2022ATel15443....1K}
{Kennea} J.~A.,  {Gronwall} C.,  {Page} K.~L.,  {Palmer} D.~M.,  {Tohuvavohu}
  A.,   {Neil Gehrels Swift Observatory Team} 2022a, The Astronomer's Telegram,
  \href {https://ui.adsabs.harvard.edu/abs/2022ATel15443....1K} {15443, 1}

\bibitem[\protect\citeauthoryear{{Kennea}, {Evans}  \& {Negoro}}{{Kennea}
  et~al.}{2022b}]{2022ATel15561....1K}
{Kennea} J.~A.,  {Evans} P.~A.,   {Negoro} H.,  2022b, The Astronomer's
  Telegram, \href {https://ui.adsabs.harvard.edu/abs/2022ATel15561....1K}
  {15561, 1}

\bibitem[\protect\citeauthoryear{{Laycock}, {Zezas}, {Hong}, {Drake}  \&
  {Antoniou}}{{Laycock} et~al.}{2010}]{2010ApJ...716.1217L}
{Laycock} S.,  {Zezas} A.,  {Hong} J.,  {Drake} J.~J.,   {Antoniou} V.,  2010,
  \mn@doi [\apj] {10.1088/0004-637X/716/2/1217}, \href
  {https://ui.adsabs.harvard.edu/abs/2010ApJ...716.1217L} {716, 1217}

\bibitem[\protect\citeauthoryear{{Leahy}}{{Leahy}}{1987}]{1987A&A...180..275L}
{Leahy} D.~A.,  1987, \aap, \href
  {https://ui.adsabs.harvard.edu/abs/1987A&A...180..275L} {180, 275}

\bibitem[\protect\citeauthoryear{{Li} \& {van den Heuvel}}{{Li} \& {van den
  Heuvel}}{1999}]{1999ApJ...513L..45L}
{Li} X.~D.,  {van den Heuvel} E.~P.~J.,  1999, \mn@doi [\apjl]
  {10.1086/311904}, \href
  {https://ui.adsabs.harvard.edu/abs/1999ApJ...513L..45L} {513, L45}

\bibitem[\protect\citeauthoryear{{Lomb}}{{Lomb}}{1976}]{1976Ap&SS..39..447L}
{Lomb} N.~R.,  1976, \mn@doi [\apss] {10.1007/BF00648343}, \href
  {https://ui.adsabs.harvard.edu/abs/1976Ap&SS..39..447L} {39, 447}

\bibitem[\protect\citeauthoryear{{Lutovinov} \& {Tsygankov}}{{Lutovinov} \&
  {Tsygankov}}{2008}]{2008AIPC.1054..191L}
{Lutovinov} A.,  {Tsygankov} S.,  2008, in {Axelsson} M.,  ed.,  American
  Institute of Physics Conference Series Vol. 1054, Cool Discs, Hot Flows: The
  Varying Faces of Accreting Compact Objects. pp 191--202 (\mn@eprint {arXiv}
  {0808.2034}), \mn@doi{10.1063/1.3002502}

\bibitem[\protect\citeauthoryear{Lutovinov, Tsygankov  \&
  Chernyakova}{Lutovinov et~al.}{2012}]{Lut2012}
Lutovinov A.,  Tsygankov S.,   Chernyakova M.,  2012, \mn@doi [Monthly Notices
  of the Royal Astronomical Society] {10.1111/j.1365-2966.2012.21036.x}, 423,
  1978

\bibitem[\protect\citeauthoryear{{Madsen} et~al.,}{{Madsen}
  et~al.}{2015}]{2015ApJS..220....8M}
{Madsen} K.~K.,  et~al., 2015, \mn@doi [\apjs] {10.1088/0067-0049/220/1/8},
  \href {https://ui.adsabs.harvard.edu/abs/2015ApJS..220....8M} {220, 8}

\bibitem[\protect\citeauthoryear{{Maitra}, {Paul}  \& {Naik}}{{Maitra}
  et~al.}{2012}]{2012MNRAS.420.2307M}
{Maitra} C.,  {Paul} B.,   {Naik} S.,  2012, \mn@doi [\mnras]
  {10.1111/j.1365-2966.2011.20196.x}, \href
  {https://ui.adsabs.harvard.edu/abs/2012MNRAS.420.2307M} {420, 2307}

\bibitem[\protect\citeauthoryear{{Mereminskiy}, {Mushtukov}, {Lutovinov},
  {Tsygankov}, {Semena}, {Molkov}  \& {Shtykovsky}}{{Mereminskiy}
  et~al.}{2022}]{2022A&A...661A..33M}
{Mereminskiy} I.~A.,  {Mushtukov} A.~A.,  {Lutovinov} A.~A.,  {Tsygankov}
  S.~S.,  {Semena} A.~N.,  {Molkov} S.~V.,   {Shtykovsky} A.~E.,  2022, \mn@doi
  [\aap] {10.1051/0004-6361/202141813}, \href
  {https://ui.adsabs.harvard.edu/abs/2022A&A...661A..33M} {661, A33}

\bibitem[\protect\citeauthoryear{{Mushtukov}, {Suleimanov}, {Tsygankov}  \&
  {Poutanen}}{{Mushtukov} et~al.}{2015a}]{2015MNRAS.447.1847M}
{Mushtukov} A.~A.,  {Suleimanov} V.~F.,  {Tsygankov} S.~S.,   {Poutanen} J.,
  2015a, \mn@doi [\mnras] {10.1093/mnras/stu2484}, \href
  {https://ui.adsabs.harvard.edu/abs/2015MNRAS.447.1847M} {447, 1847}

\bibitem[\protect\citeauthoryear{Mushtukov, Suleimanov, Tsygankov  \&
  Poutanen}{Mushtukov et~al.}{2015b}]{10.1093/mnras/stv2087}
Mushtukov A.~A.,  Suleimanov V.~F.,  Tsygankov S.~S.,   Poutanen J.,  2015b,
  \mn@doi [MNRAS] {10.1093/mnras/stv2087}, 454, 2539

\bibitem[\protect\citeauthoryear{{Naik} et~al.,}{{Naik}
  et~al.}{2008}]{2008ApJ...672..516N}
{Naik} S.,  et~al., 2008, \mn@doi [\apj] {10.1086/523295}, \href
  {https://ui.adsabs.harvard.edu/abs/2008ApJ...672..516N} {672, 516}

\bibitem[\protect\citeauthoryear{{Naik}, {Paul}, {Kachhara}  \&
  {Vadawale}}{{Naik} et~al.}{2011}]{2011MNRAS.413..241N}
{Naik} S.,  {Paul} B.,  {Kachhara} C.,   {Vadawale} S.~V.,  2011, \mn@doi
  [\mnras] {10.1111/j.1365-2966.2010.18128.x}, \href
  {https://ui.adsabs.harvard.edu/abs/2011MNRAS.413..241N} {413, 241}

\bibitem[\protect\citeauthoryear{{Naik}, {Maitra}, {Jaisawal}  \&
  {Paul}}{{Naik} et~al.}{2013}]{2013ApJ...764..158N}
{Naik} S.,  {Maitra} C.,  {Jaisawal} G.~K.,   {Paul} B.,  2013, \mn@doi [\apj]
  {10.1088/0004-637X/764/2/158}, \href
  {https://ui.adsabs.harvard.edu/abs/2013ApJ...764..158N} {764, 158}

\bibitem[\protect\citeauthoryear{{Nakajima} et~al.,}{{Nakajima}
  et~al.}{2022}]{2022ATel15453....1N}
{Nakajima} M.,  et~al., 2022, The Astronomer's Telegram, \href
  {https://ui.adsabs.harvard.edu/abs/2022ATel15453....1N} {15453, 1}

\bibitem[\protect\citeauthoryear{Okazaki, Hayasaki  \& Moritani}{Okazaki
  et~al.}{2013}]{Okazaki2013OriginOT}
Okazaki A.~T.,  Hayasaki K.,   Moritani Y.,  2013, Publications of the
  Astronomical Society of Japan, 65, 41

\bibitem[\protect\citeauthoryear{{Pavlinsky} et~al.,}{{Pavlinsky}
  et~al.}{2022}]{2022A&A...661A..38P}
{Pavlinsky} M.,  et~al., 2022, \mn@doi [\aap] {10.1051/0004-6361/202141770},
  \href {https://ui.adsabs.harvard.edu/abs/2022A&A...661A..38P} {661, A38}

\bibitem[\protect\citeauthoryear{{Pike} et~al.,}{{Pike}
  et~al.}{2022}]{2022ApJ...927..190P}
{Pike} S.~N.,  et~al., 2022, \mn@doi [\apj] {10.3847/1538-4357/ac5258}, \href
  {https://ui.adsabs.harvard.edu/abs/2022ApJ...927..190P} {927, 190}

\bibitem[\protect\citeauthoryear{{Reig}}{{Reig}}{2011}]{2011Ap&SS.332....1R}
{Reig} P.,  2011, \mn@doi [\apss] {10.1007/s10509-010-0575-8}, \href
  {https://ui.adsabs.harvard.edu/abs/2011Ap&SS.332....1R} {332, 1}

\bibitem[\protect\citeauthoryear{{Reig} \& {Nespoli}}{{Reig} \&
  {Nespoli}}{2013}]{2013A&A...551A...1R}
{Reig} P.,  {Nespoli} E.,  2013, \mn@doi [\aap] {10.1051/0004-6361/201219806},
  \href {https://ui.adsabs.harvard.edu/abs/2013A&A...551A...1R} {551, A1}

\bibitem[\protect\citeauthoryear{Reig \& Roche}{Reig \&
  Roche}{1999}]{10.1046/j.1365-8711.1999.02473.x}
Reig P.,  Roche P.,  1999, \mn@doi [Monthly Notices of the Royal Astronomical
  Society] {10.1046/j.1365-8711.1999.02473.x}, 306, 100

\bibitem[\protect\citeauthoryear{Reig, Torrejón  \& Blay}{Reig
  et~al.}{2012}]{10.1111/j.1365-2966.2012.21509.x}
Reig P.,  Torrejón J.~M.,   Blay P.,  2012, \mn@doi [Monthly Notices of the
  Royal Astronomical Society] {10.1111/j.1365-2966.2012.21509.x}, 425, 595

\bibitem[\protect\citeauthoryear{{Reig}, {Tzoubanou}  \& {Pantaulas}}{{Reig}
  et~al.}{2022}]{2022ATel15612....1R}
{Reig} P.,  {Tzoubanou} A.,   {Pantaulas} V.,  2022, The Astronomer's Telegram,
  \href {https://www.astronomerstelegram.org/?read=15612} {15612, 1}

\bibitem[\protect\citeauthoryear{{Sasaki}, {Haberl}  \& {Pietsch}}{{Sasaki}
  et~al.}{2000}]{2000A&AS..147...75S}
{Sasaki} M.,  {Haberl} F.,   {Pietsch} W.,  2000, \mn@doi [\aaps]
  {10.1051/aas:2000290}, \href
  {https://ui.adsabs.harvard.edu/abs/2000A&AS..147...75S} {147, 75}

\bibitem[\protect\citeauthoryear{{Scargle}}{{Scargle}}{1982}]{1982ApJ...263..835S}
{Scargle} J.~D.,  1982, \mn@doi [\apj] {10.1086/160554}, \href
  {https://ui.adsabs.harvard.edu/abs/1982ApJ...263..835S} {263, 835}

\bibitem[\protect\citeauthoryear{{Serino} et~al.,}{{Serino}
  et~al.}{2022}]{2022ATel15442....1S}
{Serino} M.,  et~al., 2022, The Astronomer's Telegram, \href
  {https://ui.adsabs.harvard.edu/abs/2022ATel15442....1S} {15442, 1}

\bibitem[\protect\citeauthoryear{{Shidatsu}, {Pike}, {Mihara}, {Sugizaki},
  {Negoro}, {Nakajima}  \& {Murata}}{{Shidatsu}
  et~al.}{2022}]{2022ATel15495....1S}
{Shidatsu} M.,  {Pike} S.,  {Mihara} T.,  {Sugizaki} M.,  {Negoro} H.,
  {Nakajima} M.,   {Murata} K.,  2022, The Astronomer's Telegram, \href
  {https://ui.adsabs.harvard.edu/abs/2022ATel15495....1S} {15495, 1}

\bibitem[\protect\citeauthoryear{{Tsygankov}, {Lutovinov}, {Krivonos},
  {Molkov}, {Jenke}, {Finger}  \& {Poutanen}}{{Tsygankov}
  et~al.}{2016}]{2016MNRAS.457..258T}
{Tsygankov} S.~S.,  {Lutovinov} A.~A.,  {Krivonos} R.~A.,  {Molkov} S.~V.,
  {Jenke} P.~J.,  {Finger} M.~H.,   {Poutanen} J.,  2016, \mn@doi [\mnras]
  {10.1093/mnras/stv2849}, \href
  {https://ui.adsabs.harvard.edu/abs/2016MNRAS.457..258T} {457, 258}

\bibitem[\protect\citeauthoryear{{Tsygankov}, {Mushtukov}, {Suleimanov},
  {Doroshenko}, {Abolmasov}, {Lutovinov}  \& {Poutanen}}{{Tsygankov}
  et~al.}{2017}]{2017A&A...608A..17T}
{Tsygankov} S.~S.,  {Mushtukov} A.~A.,  {Suleimanov} V.~F.,  {Doroshenko} V.,
  {Abolmasov} P.~K.,  {Lutovinov} A.~A.,   {Poutanen} J.,  2017, \mn@doi [\aap]
  {10.1051/0004-6361/201630248}, \href
  {https://ui.adsabs.harvard.edu/abs/2017A&A...608A..17T} {608, A17}

\bibitem[\protect\citeauthoryear{{Weng}, {Ge}, {Zhao}, {Wang}, {Zhang}, {Bian}
  \& {Yuan}}{{Weng} et~al.}{2017}]{2017ApJ...843...69W}
{Weng} S.-S.,  {Ge} M.-Y.,  {Zhao} H.-H.,  {Wang} W.,  {Zhang} S.-N.,  {Bian}
  W.-H.,   {Yuan} Q.-R.,  2017, \mn@doi [\apj] {10.3847/1538-4357/aa76ec},
  \href {https://ui.adsabs.harvard.edu/abs/2017ApJ...843...69W} {843, 69}

\bibitem[\protect\citeauthoryear{{Zaznobin} et~al.,}{{Zaznobin}
  et~al.}{2022}]{2022ATel15582....1Z}
{Zaznobin} I.~A.,  et~al., 2022, The Astronomer's Telegram, \href
  {https://ui.adsabs.harvard.edu/abs/2022ATel15582....1Z} {15582, 1}

\makeatother
\end{thebibliography}

\end{document}